\documentclass[aip,pof,reprint,onecolumn,amsmath,amssymb]{revtex4-1}

\usepackage{graphicx}
\usepackage{dcolumn}
\usepackage{bm}

\usepackage{subfigure}
\usepackage{amsfonts}
\usepackage{color}
\usepackage{mathtools}
\usepackage{amssymb}

\usepackage{xcolor}
\colorlet{shadecolor}{orange!15}

\usepackage{mathrsfs}
\usepackage{dsfont}
\usepackage{wasysym}

\usepackage[skins,theorems]{tcolorbox}
\tcbset{highlight math style={enhanced,
  colframe=red,colback=white,arc=0pt,boxrule=1pt}}

\newcommand{\bs}{\boldsymbol}

\usepackage{framed}
\newcommand{\ds}{\displaystyle}
\newcommand{\beq}{\begin{equation}}
\newcommand{\eeq}{\end{equation}}
\newcommand{\beqa}{\begin{eqnarray}}
\newcommand{\eeqa}{\end{eqnarray}}

\newcommand{\bem}{\begin{math}}
\newcommand{\eem}{\end{math}}
\newcommand{\rar}{{\rightarrow}}

\newcommand{\bfr}{{\bf r}}

\newcommand{\bfJ}{{\bs J}}

\def\strutdepth{\dp\strutbox}
\def\nw#1{\strut\vadjust{\kern-\strutdepth\vtop to0pt{\vss\hbox to\hsize
{\hskip\hsize\hskip5pt$\leftarrow$\hss\strut}}}{\em #1}}
\usepackage{multirow}
\def\tl#1{\textcolor{black}{#1}}
\def\tlc#1{\textcolor{purple}{\it #1}}
\def\tll#1{\textcolor{black}{#1}}
\def\tbl#1{\textcolor{black}{#1}}
\def\yhy#1{\textcolor{black}{#1}}

\begin{document}


\title[]{Shape dependent phoretic propulsion of slender active particles}

\author{Y. Ibrahim}
%
 \affiliation{School of Mathematics, University of Bristol, Bristol BS8 1TW, UK}
 \author{R. Golestanian}
 \email{ramin.golestanian@physics.oxford.ac.uk}
\affiliation{Rudolf Peierls Centre for Theoretical Physics, University of Oxford, Oxford OX1 3NP, UK}
\author{T. B. Liverpool}%
 \email{t.liverpool@bristol.ac.uk}
\affiliation{School of Mathematics, University of Bristol, Bristol BS8 1TW, UK} 
\affiliation{BrisSynBio,  Life Sciences Building, University of Bristol, Bristol BS8 1TQ, UK 
}%


\date{\today}

\begin{abstract}
\tbl{We theoretically study the self-propulsion of a thin (slender) colloid driven by asymmetric chemical reactions on its surface at vanishing Reynolds number. Using the method of matched asymptotic expansions, we obtain the colloid self-propulsion velocity as a function of its shape and surface physico-chemical properties. The mechanics of self-phoresis for rod-like swimmers has a richer spectrum of behaviours than spherical swimmers due to the presence of two small length scales, the slenderness of the rod and  the width of the slip layer. This leads to subtleties in taking the limit of vanishing slenderness. As a result, even for very thin rods, the distribution of curvature along the surface of the swimmer, namely its shape, plays a surprising role in determining the efficiency of propulsion. We find that thin cylindrical self-phoretic swimmers with {\em blunt} ends move faster than thin prolate spheroid shaped swimmers with the same aspect ratio. }

%
\end{abstract}

\pacs{pacs}
\keywords{keywords}
\maketitle

\section{\label{sec:level1} Introduction}
%
%
%
In recent years there has been an explosion of interest in the area of soft active  materials~\cite{Marchetti2013,ramaswamy2010,Julicher2007}. Active materials are condensed matter systems driven out of equilibrium by components that autonomously convert stored energy into motion. These materials are of interest partly because of their potential as a basis for engineered smart materials. They may in addition serve as model systems for understanding fundamental non-equilibrium phenomena in cellular biology. While one the major foci of the field involves the emergence of collective behaviour from active components~\cite{Marchetti2013,ramaswamy2010,Julicher2007}, it is of particular importance to first understand and control the behaviour of the individual components that drive the system out of equilibrium~\cite{golestanian2005propulsion,yariv2010electrokinetic,michelin2014phoretic,paxton2005motility,yariv2013electrophoretic,ebbens2014electrokinetic}. 
A commonly studied active component is a synthetic self-phoretic swimmer\cite{golestanian2005propulsion,yariv2010electrokinetic,michelin2014phoretic,paxton2005motility,yariv2013electrophoretic,ebbens2014electrokinetic}. These are colloidal particles with an asymmetric catalytic coating over their surface suspended in a viscous fluid in which are dissolved chemically reactive species (fuel) whose consumption (or production) is accelerated by the catalyst\cite{golestanian2007designing}. The asymmetric catalytic coating leads to asymmetric adsorbtion and release of the reactants and products. This asymmetry of the  distribution of reactants and/or products coupled with a short-range interaction of the chemical species with the colloid surface,  leads to an interfacial flow and hence to motion of the colloid~\cite{anderson01,anderson1982motion}. 
Experimentally, the two most studied self-phoretic swimmer shapes are asymmetrically coated spheres and rods.
Among the earliest studied synthetic self-phoretic swimmers were cylindrical bi-metallic nano-rods of micron length and nm-scale diameter (with two-halves made of different metal catalysts) \cite{paxton2004catalytic,posner01} which speed up the decomposition of hydrogen peroxide (H$_2$O$_2$), in a H$_2$O$_2$ solution. Other well studied system are spherical colloids partially coated with catalysts~\cite{Howse2007,palacci2013living} also in hydrogen peroxide solutions.
Unravelling the collective behaviour of these systems~\cite{Yoshinaga2017,Saha2014} in order to design and control their macroscopic behaviour\cite{theurkauff2012dynamic,palacci2013living} requires a detailed understanding of the  mechanisms of swimmer propulsion at the single particle level. 
Given this there have been several theoretical studies aimed at isolating the physical mechanisms behind the observed propulsion and linking them 
to properties of the swimmers such as their shape and coating~\cite{golestanian2007designing,paxton2005motility,Moran2010,popescu2010phoretic,Sabass2012,schnitzer2015osmotic,jewell2016catalytically,nourhani2016geometrical}. These are all based on generalising results for phoretic motion in external gradients~\cite{anderson01} to the situation where the gradient is self-generated.

In this article we focus on calculating the propulsion speed of self-phoretic rods as a function of their slenderness ratio (defined as the ratio of the cross-sectional diameter to the rod length), $\epsilon$. 
\tbl{For thin rods with vanishing aspect ratio, $\epsilon$  there is some debate about the particular form of the dependence on the self-propulsion speed on the length and aspect ratio~\cite{golestanian2007designing,schnitzer2015osmotic,nourhani2016geometrical}. Here we show this is because there are two small length scales in the problem, (1) the diameter of the rod and (2) the thickness of the phoretic slip layer. By a careful  asymptotic analysis, we identify some subtleties in taking the limits of both these length scales to zero which clarifies the debate.
We point out that physically, there is a clear hierarchy of length scales which implies that the thickness of the slip layer (a molecular scale)  is much smaller than rod diameter (a colloidal scale) which is much smaller than the rod length. When we take this hierarchy into account, we find that details of the shape of the rods becomes of paramount importance in determining the asymptotic propulsion speed (in the limit $\epsilon \rar 0$). 
It has been widely assumed that thin cylindrical self-phoretic rods studied experimentally can be approximated by an equivalent thin spheroid (ellipsoid of revolution) of the same slenderness ratio\cite{yariv2010electrokinetic,schnitzer2015osmotic,nourhani2016geometrical}. Here we examine this assumption by comparing slender active particles of fixed slenderness $\epsilon$ whose shape are given by thin spheroids with those whose shape is given by cylindrical rods and obtain asymptotic results in the limit  $\epsilon \rar 0$. One conclusion of our analysis is that in fact, for issues of self-propulsion, this assumption is  in general, not  a justified one.
} 
 
 \par To be concrete, we do this by using the method of matched asymptotic expansions for swimmers with a generic shape function that interpolates between needle-like spheroid and cylindrical rod-like geometries.  We find that rod-like shapes are faster than the needle-like shapes with the same $\epsilon$. Our results thus suggest a that a strategy for better performance (faster speed) of self-phoretic swimmers would be to  design blunt rod-like shapes rather than tapered  needle-like shapes.

\begin{figure}[h!]
\begin{center}
\includegraphics[scale=.4]{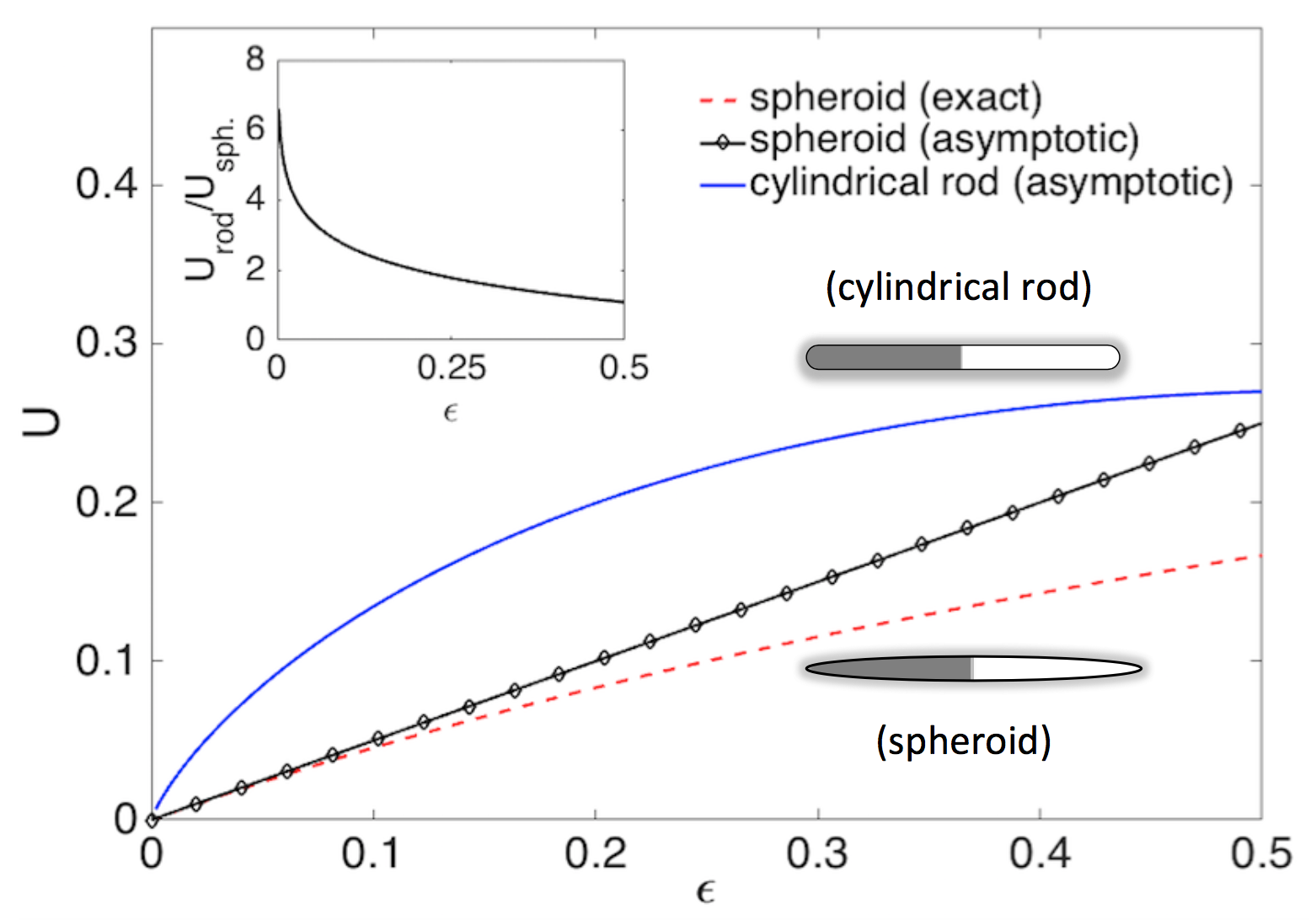}

%
\caption{Comparison of swimmer propulsion speeds.  The propulsion speed, $U$ of a half-coated rod as a function of the slenderness ratio, $\epsilon$. The speed, $U$, is expressed in units of \yhy{$|\mu^*\alpha^*/D|$}. The blue solid line is the asymptotic result ($\epsilon \rar 0$) for the cylindrical rod (see eqn. (\ref{rod:asymptotic:speed})), while the black diamonds indicate the asymptotic result for the spheroid (see eqn. (\ref{spheroid:asymptotic:speed}) and ref.\cite{schnitzer2015osmotic}).  The dashed red line is the plot of the exact propulsion speed obtained in ref. \cite{nourhani2016geometrical} of a half-coated spheroid. (inset) Plot of the ratio of the asymptotic speeds of the cylindrical rod to the spheroid asymptotic against the slenderness ratio (see eqns. (\ref{spheroid:asymptotic:speed}) and (\ref{rod:asymptotic:speed})).  }
\end{center}
\end{figure}
\tl{
\section{
The thin diffusiophoretic swimmer}}
\tl{
We consider a thin (slender) colloid which has 
a heteregeneous  
catalytic surface that produces (or consumes) molecules of a chemical species in a variable manner on its surface generating a local concentration gradient of the species in its vicinity. 
We then use the standard framework~\cite{anderson01, golestanian2007designing} for studying phoretic propulsion to calculate the steady-state speed of the colloid generated by this self-generated gradient. This is done by solving the coupled equations of motion of the local concentration of reactants/products and local momentum conservation taking account of the boundary conditions on the colloid surface.
}
\\
\begin{figure}[h!]
\begin{center}
\subfigure[]{
\includegraphics[scale=.28]{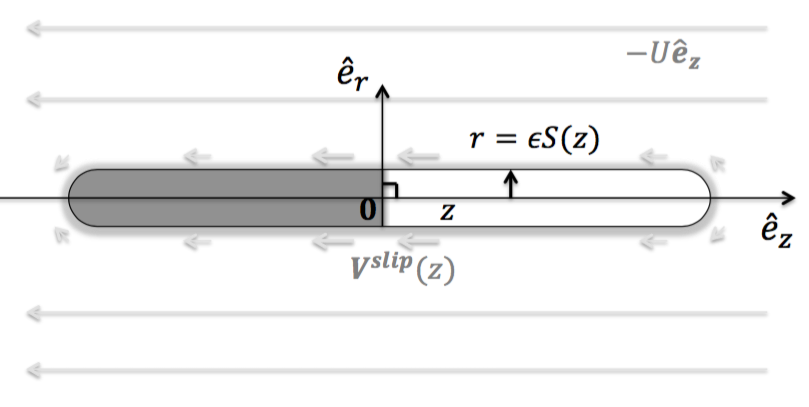}
}
\subfigure[]{
\includegraphics[scale=.32]{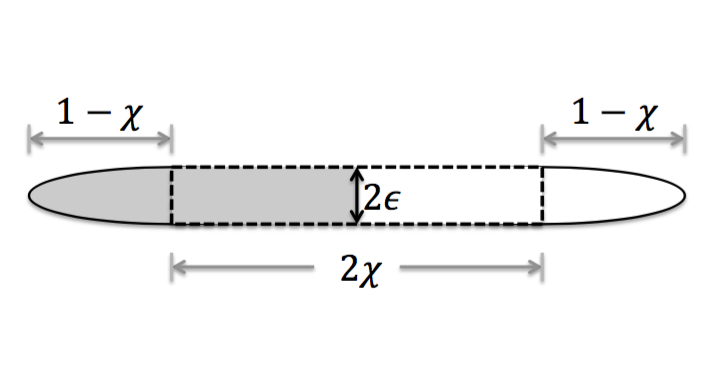} \label{generic:shape}
}  
\caption{\tl{(a) Sketch of a  slender catalytic microswimmer showing the variation of catalytic coverage and shape along the long axis oriented in the  $z$ direction. The gray (shaded) half of the swimmer surface is made of a different catalyst than the unshaded half. The catalysts produce (or consume)  molecules of the reactive species at different rates leading to an asymmetric distribution of molecules near the swimmer. $\bs{V}^{\mbox{slip}}$ is the phoretic slip velocity induced by the coupling of the  asymmetric distribution of the  molecules and their short-range interaction with the swimmer surface. $U \hat{\bs{e}}_z$ is the axial translational velocity of the swimmer in the lab frame. (b) The generic thin rod with rounded ends which interpolates between a thin spheroid  and a thin cylinder. The limit $({\chi} = 0)$ is the spheroid, while $( {\chi} = 1)$ is the cylindrical rod.}}
\end{center}
\end{figure}

\subsection{Equations of motion} 
\tl{
\emph{Concentration}: 
The  concentration field of the reactive species $\widetilde{c}(\bfr)$ obeys in the steady state,
\begin{equation}
\widetilde{\nabla} \cdot \widetilde{\bs{J}}  = 0; \qquad \widetilde{\bs{J}} = - D \left( \widetilde{\nabla} \widetilde{c}  + \frac{ \widetilde{c} }{k_BT} \widetilde{\nabla} \widetilde{\psi} \right)  \   ,  
\end{equation}
in the zero P\'eclet number limit, where $\widetilde{\psi}(\widetilde{r} - a S(\widetilde{z}),\widetilde{z})$ is the short-range interaction energy of the reactants (products) with the colloid surface. $S(\widetilde{z})$ is the colloid shape function, the surface of the colloid is given by the function $\widetilde{\bfr} (\widetilde{z},\theta) = (a S(\widetilde{z}) \cos \theta ,  a S(\widetilde{z}) \sin \theta ,\widetilde{z})$. In the experimental systems motivating this work, the typical length of the self-phoretic colloids is $2l \sim 2 \mu$m and they move with a speed $U^* \sim 10 \mu$ms$^{-1}$ while the diffusion coefficient of the reactants (products) $D \sim 10^3 \mu $m$^2$s$^{-1}$. Therefore, they are in the small P\'eclet number limit ($\mathcal{P}e = U^*l /D \ll 1$). $k_B$ is the Boltzmann constant and $T$ is the temperature of the solution.
The heterogeneous surface leads to a variable production/consumption rate, $\widetilde{\alpha}(\widetilde{z})$ of molecules on the surface giving  a
flux boundary condition on the surface: 
\begin{align}
& \bs{\hat{n}} \cdot \widetilde{\bs{J}}  = \widetilde{\alpha}(z), \quad \mbox{at} \quad \widetilde{r} = a \ S(\widetilde{z}),
\end{align}
and the concentration attenuation decays far from the swimmer, $\widetilde{c} \rightarrow \widetilde{c}_{\infty},$ as  $ \sqrt{\widetilde{r}^2 + \widetilde{z}^2} \rightarrow \infty$.   \\
We restrict our analysis to the fixed flux limit where the catalytic production/consumption rate, $\alpha(z)$, does not depend on the local concentration of the reactant (product) particles. In this regime, the particles flux is limited by the reaction rate rather than diffusion of the particles to the swimmer surface. }
\par \tl{\emph{Momentum conservation}: 
The fluid velocity $\widetilde{\bs{v}}(\bfr)$, and hydrostatic pressure $\widetilde{p}(\bfr)$, obey the Stokes' equations of vanishing Reynolds number incompressible flow
\begin{align}
 0  & = \widetilde{\nabla} \cdot \widetilde{\bs{v}}  \ ,  \\
 \mathbf{0} & = \eta \widetilde{\nabla}^2 \widetilde{\bs{v}} - \widetilde{\nabla} \widetilde{p} -  \widetilde{c} \  \widetilde{ \nabla} \widetilde{\psi}  \  , 
\end{align}
where $\eta$ is the fluid viscosity. This is justified as typical experimental swimming speeds of $U^* \sim 10\mu$ms$^{-1}$ and $2l \sim 2\mu$m colloid length, and for typical solution kinematic viscosity $\nu\sim 10^{-6}$m$^2$s$^{-1}$, mean these colloids generate flows  with very low Reynolds number ($\mathcal{R}e = U^* l/\nu \ll 1$). The inhomogeneous term $-  \widetilde{c} \nabla \widetilde{\psi}$ arises due to the interactions of the reactant (product) particles with the colloid surface, $\widetilde{\psi}(r- a S(z),z)$. We have no-slip boundary conditions on the colloid surface in otherwise uniform flow 
\begin{align}
\widetilde{\bs{v}} & = \bs{0},  \quad  \mbox{at} \quad \widetilde{r} = a \ S(\widetilde{z}), \\
\widetilde{\bs{v}} & \rightarrow - \widetilde{U} \bs{\hat{e}}_z, \quad \sqrt{\widetilde{r}^2 + \widetilde{z}^2} \rightarrow \infty.
\end{align}
Note that we are concerned only with the axial propulsion of the swimmer (restricting the swimmer translation along $\widetilde{z}$-axis).
In addition, the colloid 
experiences no net body force (which uniquely determines $\widetilde{U}$)
\begin{equation}
\oiint_{\widetilde{r} = a S(\widetilde{z})} \widetilde{\bs{\Pi}} \cdot \bs{\hat{n}} \ d \widetilde{\mathcal{S}} -  \iiint_{-\infty}^{\infty} \widetilde{c} \,  \widetilde{\nabla} \, \widetilde{\psi} \ d\widetilde{V} = \mathbf{0} \label{force:free}
\end{equation} 
where $\widetilde{\bs{\Pi}} = - \widetilde{p} \mathds{1} + \eta \left( \widetilde{\nabla}  \, \widetilde{\bs{v}} + \left[ \widetilde{\nabla} \,  \widetilde{\bs{v}} \right]^T \right)$ is the fluid stress tensor and $d \widetilde{\mathcal{S}}=a d \phi d z$, $d\widetilde{V}=r dr d\phi dz$ are the surface and volume elements in cylindrical coordinates respectively.
}\\

\subsection{Non-dimensionalization}
\tl{We express the displacements $(r,z)$ in terms of half the length of the swimmer, $l$. We consider swimmers with  concentration gradients generated by the colloids on a  scale $c^*/a$. Hence we 
 express the  flux $\widetilde{J}$ of reactant (product) particles in units of  $c^*D / a$, the concentration field $\widetilde{c}$ in units of $c^*$,  (where $D$ is the diffusion coefficient of the reactant/product particles), the interaction energy $\widetilde{\Psi}$ in the units of thermal energy $k_BT$, ($k_B$ the Boltzmann constant and $T$ is the temperature of the solution), the flow field $\widetilde{\bs{v}}$ with $U^* =  \mu^* c^*/a$ (where $\mu^* =  k_BT L^{*2} /\eta$ is the characteristic phoretic mobility coefficient, $\eta$ is the viscosity of the fluid and $L^{*}$ is the short-range interaction lengthscale of the reactant (product) molecules with the swimmer surface).  We scale the pressure field $\widetilde{p}$ with $\eta U^*/l$. 
We note that the problem has three length-scales; the swimmer characteristic length $2l$, its cross-sectional radius $a$ and the range $L^*$ of the interaction between the molecules and swimmer surface. Consequently, we have three asymptotic near-field regions and in addition the scale on which the  ends of the rod are rounded. }

\tl{
We therefore define the dimensionless parameters, 
 $\epsilon=a/l$, the slenderness ratio and $\lambda=L^*/l$, the interaction layer thickness to the swimmer largest lengthscale, with $0 < \lambda \ll \epsilon \ll 1$. In addition, we define dimensionless fields 
\begin{math}
\bs{J}  = \bs{\widetilde{J}} a / c^* D, \;
c  = \widetilde{c}/c^*, \;
\bs{v}  = \widetilde{\bs{v}}/U^*, \;
p  = \widetilde{p} l/\eta U^*, \;
\psi  = \widetilde{\psi} / k_BT,
\end{math}
and the dimensionless catalytic flux $\alpha  = \widetilde{\alpha} a /c^* D$ on the swimmer surface.
}
%
%
%
%

\tl{
Hence we obtain dimensionless equations of motion, 
\begin{equation}
\nabla \cdot \bs{J}  = 0\; ; \qquad \bs{J} = - \epsilon \left( \nabla c + c \nabla \psi \right)  \  ,   \label{gov:mass:transfer}
\end{equation}
\begin{align}
 0  & = \nabla \cdot \bs{v} \  ,  \label{gov:continuity}  \\
 \mathbf{0} & = \nabla^2 \bs{v} - \nabla p - \epsilon \ \lambda^{-2} \ c \nabla \psi  \  ,  \label{gov:stokes}
 \end{align}
 %
 %
 %
 %
where $\lambda = L^{*}/l$ is the ratio of the interaction length-scale to half the length of the swimmer, $\epsilon = a/l$ is the swimmer slenderness ratio and $\psi(r-\epsilon S(z),z)$ is the short-range interaction  potential between reactant (product) molecules and surface.
}
\tl{ The fixed flux boundary condition of the chemical species at the colloid's surface is now 
\begin{align}
& \bs{\hat{n}} \cdot \bs{J}  = \alpha(z), \quad \mbox{at} \quad r = \epsilon \ S(z),
\end{align}
and the  concentration  decays to its value far from the swimmer, $c \rightarrow c_{\infty}, \ \sqrt{r^2 + z^2} \rightarrow \infty$.  
}
\tl{
 The flow field  boundary conditions are now
\begin{align}
\bs{v} & = \bs{0},  \quad  \mbox{at} \quad r = \epsilon \ S(z), \\
\bs{v} & \rightarrow - U \bs{\hat{e}}_z, \quad \sqrt{r^2 + z^2} \rightarrow \infty.
\end{align}
%
%
The zero torque and force conditions are 
\begin{equation}
\oiint_{r = \epsilon \, S(z)} \bs{\Pi} \cdot \bs{\hat{n}} \ d \mathcal{S} - \epsilon \ \lambda^{-2} \iiint_{-\infty}^{\infty} c \  \nabla \psi \ dV = \mathbf{0} \label{force-t:free}
\end{equation} 
where $\bs{\Pi} = - p \mathds{1} +  \left( \nabla \bs{v} + \left[ \nabla \bs{v} \right]^T \right)$ is the dimensionless stress tensor and $d \mathcal{S}$, $dV$ are the surface and volume elements in cylindrical coordinates respectively.
}
\subsection{The slender shape function}
\tl{
For the analysis in this paper, we consider a generic shape function of a cylindrical rod with hemispheroidal caps (see Figure \ref{generic:shape}) 
\begin{equation}
S(z; {\chi} ) =
\begin{dcases}
1, & 0 \leq |z| \leq  {\chi}; \\
\quad \\
 \sqrt{1 - \left(\frac{|z| - {\chi}}{1-{\chi}} \right)^2 }, &  {\chi} \leq  |z|  \leq 1, 
\end{dcases}  \label{shape:function}
\end{equation}
where the parameter $0 \leq |{\chi}| < 1$ is the additional lengthscale specifying the swimmer shape curvature. The limit ${\chi} \rightarrow 0$ is the widely studied spheroid shape, while the limit ${\chi} \rightarrow 1$ is the cylindrical rod shape. $1-{\chi} = \epsilon$ is a cylindrical rod with hemispherical caps.
}
\par \tl{The vector function, $\bs{x}(z,\theta) = \epsilon S(z) \bs{\hat{e}}_r (\theta) + z \bs{\hat{e}}_z$, ($-1 \leq z \leq 1$), describes the local axisymmetric swimmer surface (i.e straight backbone with circular cross-section). Therefore, the unit tangent and normal vectors to the surface  are 
\begin{align}
\bs{\hat{t}} & = \frac{ \epsilon S_z(z) \bs{\hat{e}}_r + \bs{\hat{e}}_z }{\sqrt{1 + \epsilon^2 S_z^2(z)}}, \label{unit:vector:tangent} \\
\bs{\hat{n}} & = \frac{\bs{\hat{e}}_r  - \epsilon S_z(z) \bs{\hat{e}}_z }{\sqrt{1 + \epsilon^2 S_z^2(z)}},  \label{unit:vector:normal}
\end{align}
where $S_z(z) = dS(z)/dz$. 
}
\section{Analysis}
\begin{figure}[h!]
\begin{center}
\includegraphics[scale=.35]{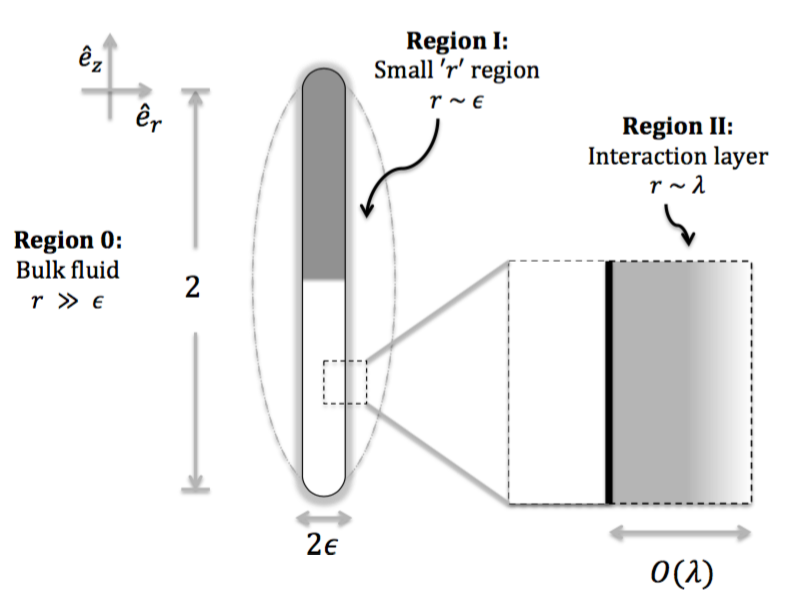}
\caption{\yhy{Sketch of the slender swimmer showing the asymptotic regions. $\lambda = L^*/l$ and $\epsilon = a/l$ are dimensionless small parameters; where $2l$ is the swimmer length, $2a$ its diameter and $L^*$ the range of the interaction between  reactant (product) molecules and the swimmer surface (typically of the order of molecular size $\sim \AA$).}}  \label{fig:slender:asymp:regions}
\end{center}
\end{figure}

%
\tl{\subsection{Overview}
We use the method of matched asymptotics together with slender-body approximations to solve for the swimmer propulsion speed $U$. We shall first identify the different regions of the solution of the equations of motion, and define re-scaled coordinates for both the interaction layer $r \sim \mathcal{O}(\lambda)$ and the slender body inner region $r \sim \mathcal{O}(\epsilon)$ \yhy{ (see Figure \ref{fig:slender:asymp:regions})}. In the subsequent sections, we systematically find solutions to the diffusion and Stokes equations in these regions. Then we  match these asymptotic results to obtain a consistent solution for the whole domain and hence find the propulsion speed $U$ which is the goal of the analysis.
}

%
%
\tll{We have three regions of solution for our matched asymptotics. In the outer layer, region 0, $ r \gg \epsilon,\lambda $; in the intermediate layer, region I, $ r \sim \epsilon$; and in the inner interaction layer, region II, $ r \sim \lambda$. In regions 0 and I, the hierarchy of scales $\epsilon \gg \lambda$ mean that we can approximate the interaction layer as being of vanishing width: we can set $\lambda$ to zero, and our analysis proceeds via an asymptotic expansion purely in $\epsilon$. However in region II, we must take account of both $\lambda,\epsilon$ and need to perform a {\em double expansion} in both variables which we use for our asymptotic analysis. 
The details of our calculation are outlined in the appendix and we present only a summary of the main results here.}

\tl{As one moves further away from the swimmer  (i.e $ r \gg \epsilon $), it tends to a line of vanishing thickness of length $2$, allowing us to use the slender body approximation  in the outer region~\cite{Batchelor70Slender,cox1970motion,keller1976slender,geer1976stokes}, to calculate the fluid velocity, $\bs{v}(r,z)$ in the direction parallel to the swimmer axis (the axial component of the velocity), $v_{z}(0,z) = \hat{\bs{e}}_z \cdot  \bs{v}(0,z)$, 
\begin{equation}
U +   \left( \hat{\bs{e}}_z \cdot \hat{\bs{t}}(z) \right) v_{z}(0,z) \ =  4 q(z;\epsilon)  \log{\left( \frac{1}{\epsilon} \right)} + 4 q(z;\epsilon) \log\left( \frac{2 \sqrt{1-z^2} }{S(z)} \right) - 2q(z;\epsilon) + 2 \int_{-1}^{1}  \frac{q({\xi};\epsilon)  - q(z;\epsilon)}{|z - {\xi}|} d{\xi} \  ,  \label{summary:speed:axial:matching}
\end{equation}   }
where the slip velocity, $\bs{v}(0,z) =\bs{V}^{\text{slip}}$ and force density along the swimmer centerline, $q(z,\epsilon)$ are obtained by matching with the inner and intermediate regions (see Appendix \ref{app:reg0}) . 

The propulsion speed $U$ is determined by requiring that the solution of the  integral equation above satisfies the constraint of zero total force on the swimmer, i.e requiring
\beq \int_{-1}^{1}q(z)dz=0 \quad . \label{eq:summary:zero_force}\eeq
%
%
 
%
%
From region I and II, we obtain the slip velocity~\cite{anderson01} as (see Appendices \ref{app:regI} and \ref{app:regII})
\begin{equation}
  \bs {V}^{\text{slip}}(z;\epsilon) \  =  \     \epsilon \ \mu(z) \  \hat{\bs{t}} \cdot \nabla c (z;\epsilon)  \  \hat{\bs{t}} \    \  ,  \label{summary:slip:velocity}
\end{equation}
where we have define the diffusiophoretic mobility~\cite{anderson01} $\mu(z) = \int_0^\infty \ {\rho} \left[ 1  - e^{- \Psi^{(0)}({\rho},z) } \right] \mathrm{d}{\rho}$, and the concentration gradient, $\nabla c $ is obtained from region 0 and I.

The concentration, $c (z;\epsilon)$  in region 0 and I can be calculated as (see Appendices \ref{app:reg0} and \ref{app:regI})
\begin{equation}
c (z;\epsilon) =  C_I^{(-1)} (z) \log \left( \frac{1}{\epsilon}\right)  + C_I^{(0)}(z)  
\label{summary:solute:field:intermediate:matching}
\end{equation}
where 
\begin{align}
C_I^{(-1)}(z) & = \alpha(z) S(z)  \  ,  \label{summary:solute:expansion:1}  \\
C_I^{(0)}(z)  & =
  C_{\infty} +  \alpha(z) S(z) \log \left( \frac{2}{S(z)} \right) - \frac{1}{2} \int_{-1}^1 \frac{\mathrm{d}}{\mathrm{d} \xi } \Big[\alpha({\xi})S({\xi})\Big] \ \mbox{sgn}\left( z - {\xi} \right) \log|z-{\xi}| \ d{\xi}  \  .
 \label{summary:solute:expansion:0}
\end{align}  \\

\par \tl{So we can calculate the propulsion speed of the swimmer by solving the coupled integral equations (\ref{summary:speed:axial:matching},\ref{eq:summary:zero_force}) for the unknowns $q(z),U$. The solutions will depend on the shape functions $S(z)$ due to the explicit dependence on $S(z)$ of equation (\ref{summary:speed:axial:matching}) and the fact that  $\hat{\bs{t}}(z) = \hat{\bs{t}} (S(z))$ . In the next section, we will consider slender swimmers with both needle-like (or spheroid) and rod-like shapes and highlight the different asymptotic limits of their propulsion velocities. }

\subsection{The swimmer propulsion velocity\label{propulsion:speed:calcs}}

\tl{In this section, we obtain an approximate value for the propulsion speed $U$ of the swimmer and  force density, $q(z;\epsilon)$ by solving the coupled integral equations (\ref{summary:speed:axial:matching},\ref{eq:summary:zero_force}). There are various methods for solving this type of problem. For example, an approximate solution by a Legendre expansion of the force density\cite{tuck1964some} $q(z;\epsilon)$ in the interval $-1 \leq z \leq 1$ to obtain an infinite linear system of equations from which one obtains $U$. Other alternatives are  perturbative methods\cite{keller1976slender} for solving integral equations where the force densities (and hence the propulsion speed $U$) are found iteratively. However, since we are interested here in the slender limit, following Cox~\cite{cox1970motion} and Tillett~\cite{Tillett1970}, the presence of $\log (\epsilon)$ in equation (\ref{summary:speed:axial:matching}) suggests an asymptotic series expansion of $q(z)$ and $U$ with terms of type $\epsilon^m (\log (\epsilon))^n$, with $n,m$ integers. Since the slip velocity vanishes as $\epsilon \; \rar \; 0$, in the limit 
  $\epsilon \ll 1$, the asymptotically leading terms in such a series will be of the form $\epsilon (\log \epsilon)^n$ where $n \le 1$, 
  which suggests an expansion~\cite{cox1970motion,Tillett1970,schnitzer2015osmotic} of the force density and speed in powers of $\left(\log\left(\epsilon\right)\right)^{-1}$.
} 

%
\tl{From equations (\ref{summary:slip:velocity}) and (\ref{summary:solute:field:intermediate:matching}) , we can write 
\begin{equation}
V_z^{\text{slip}}(z;\epsilon) \  =  \  V_{-1}^{\text{slip}}(z) \ \epsilon \log \left( \frac{1}{\epsilon}\right) \    \left( \hat{\bs{e}}_z \cdot \hat{\bs{t}} \right)  \ + \  V_0^{\text{slip}}(z) \ \epsilon \   \left( \hat{\bs{e}}_z \cdot \hat{\bs{t}} \right)  \  ,  \label{slip:expansion}
\end{equation}
where
\begin{align*}
V_{-1}^{\text{slip}}(z) & =   \mu(z) \ \hat{\bs{t}} \cdot \nabla \  C_I^{(-1)} (z) \quad  \ , \quad 
V_0^{\text{slip}}(z)  =  \mu(z) \ \hat{\bs{t}} \cdot \nabla \  C_I^{(0)}(z)  \  .
\end{align*}   
with $C_I^{(-1)},C_I^{(0)}$ defined in equations (\ref{summary:solute:expansion:1},\ref{summary:solute:expansion:0}).}
The propulsion speed $U$ and the force density $q(z;\epsilon)$ are written as expansions in $(\log \epsilon)^{-1}$
\begin{align}
{U  \over \epsilon} & = - U_{-1}  \log \left( {\epsilon} \right)  + U_0  -  U_1  \left( \log \left( {\epsilon} \right) \right)^{-1} + \cdots  \  , \label{U:expansion} \\
  { q(z;\epsilon) \over \epsilon}  & = q_0 (z) - q_1(z)  \left( \log \left( {\epsilon} \right) \right)^{-1} - q_2 (z)  \left( \log \left( {\epsilon} \right) \right)^{-2} + \cdots   \   ,  \label{q:expansion}
\end{align}
\tll{where the force-free condition and the validity of the expansion for all values of $\epsilon$, implies, $\int_{-1}^1 q_n (z) \ dz = 0$, for all $n$.} Substituting equations (\ref{U:expansion},\ref{q:expansion}) and equation (\ref{slip:expansion})  into equation (\ref{summary:speed:axial:matching}) gives 
\begin{align}
 U_{-1} + V_{-1}^{\text{slip}}(z) \ \left( \hat{\bs{e}}_z \cdot \hat{\bs{t}} \right)  & \  =  \  4q_0(z) \ ,  \label{U:_1} \\
 U_0 +  V_0^{\text{slip}}(z) \ \left( \hat{\bs{e}}_z \cdot \hat{\bs{t}} \right) & \  =  \    4 q_0(z;\epsilon) \log\left( \frac{2 \sqrt{1-z^2} }{S(z)} \right) - 2q_0(z;\epsilon) + 2 \int_{-1}^{1}  \frac{q_0({\xi};\epsilon)  - q_0(z;\epsilon)}{|z - {\xi}|} \mathrm{d}{\xi}  \ ,  \label{U:0} \\
U_n &  \   =  \    4 q_n(z;\epsilon) \log\left( \frac{2 \sqrt{1-z^2} }{S(z)} \right) - 2q_n(z;\epsilon) + 2 \int_{-1}^{1}  \frac{q_n({\xi};\epsilon)  - q_n(z;\epsilon)}{|z - {\xi}|} \mathrm{d}{\xi}   \ . \label{U:n:all}
\end{align}
\par \tl{So applying the force-free condition, we get a general asymptotic expression for the speed,
\beq
U= U_{-1} \epsilon \log(1/\epsilon) + U_0 \epsilon  + \mathcal{O}\left(\epsilon \left(\log(1/\epsilon)\right)^{-1}\right)
\quad .
\eeq
where the leading contribution to the speed (from equation (\ref{U:_1})) is 
\begin{equation}
U_{-1} = - \frac{1}{2} \int_{-1}^{1}  \left( \hat{\bs{e}}_z \cdot \hat{\bs{t}} (z) \right) \  V_{-1}^{\text{slip}}(z) \ \mathrm{d}z   
\  ,
\end{equation} 
%
and the next  term} 
 (from \tll{equation (\ref{U:0})) is 
\begin{equation}
U_0 = - \frac{1}{2} \int_{-1}^{1}  \left( \hat{\bs{e}}_z \cdot \hat{\bs{t}} (z)  \right) \  V_0^{\text{slip}}(z) \ dz -  2 \int_{-1}^1  q_0(z) \log{\left( \frac{2 \sqrt{1-z^2}}{S(z)} \right)}  \ \mathrm{d}z  \ ,
\end{equation}
where
\beq
q_0(z) =  \frac{1}{4} \left( \hat{\bs{e}}_z \cdot \hat{\bs{t}} \right)  V_{-1}^{\text{slip}}(z)   -  \frac{1}{8} \int_{-1}^1 \left( \hat{\bs{e}}_z \cdot \hat{\bs{t}} \right)  V_{-1}^{\text{slip}}(z) \,  dz  
 \quad .
\eeq
Using equation (\ref{U:n:all}) all the higher order contributions to $U$ can in principle be evaluated. 
However, if one intends to go beyond asymptotic results 
iterative methods might converge faster than the expansion outlined above~\cite{keller1976slender}. We note that for a spheroidal rod where $S(z)=\sqrt{1-z^2}$, we can explicitly evaluate the integral $\int_{-1}^1 q_n (z) \ dz = 0$ in equation (\ref{U:n:all}) and one finds that all $U_n=0, n \ge 1$. \\
%
}

\par
\begin{figure}[h!]
\begin{center}
\subfigure[\quad ${\chi} = 0$]{
\includegraphics[scale=.22]{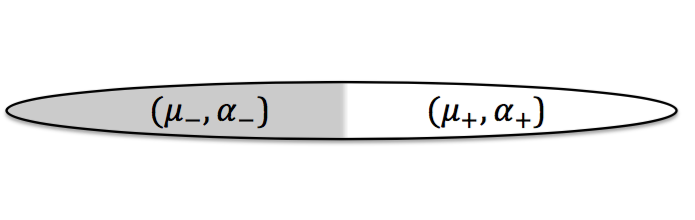} \label{eg:shape:spheroid}
} 
\subfigure[\quad $0 \leq {\chi} \leq 1$]{
\includegraphics[scale=.22]{rod_spheroid_limits} \label{eg:rod:spheroid}
}
\subfigure[\quad ${\chi} = 1$]{
\includegraphics[scale=.22]{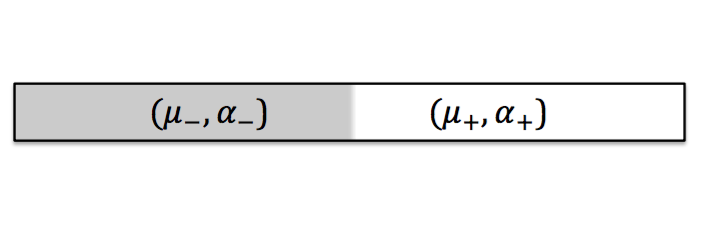} \label{eg:shape:rod}
} 
%
%
\caption{Examples of slender self-phoretic swimmer shapes with different physico-chemical properties on their surfaces.} \label{shape:examples}
\end{center}
\end{figure}

 \tl{Now, we will use these results to calculate the propulsion speed of slender swimmers, with a variety of shapes, $S(z)$ and surface activities, $\alpha(z)$.  We consider swimmers in which the activity,  $\alpha(z)=\alpha_{+}$ on the right segment $0<z \leq 1$ and $\alpha(z)=\alpha_{-}$ on the left segment $-1 \leq z < 0$ and with uniform mobility $\mu(z) = \mu_{+} = \mu_{-} = 1$ (see Figure  \ref{eg:shape:spheroid}, \ref{eg:rod:spheroid}, \ref{eg:shape:rod}). In most of our analysis we use a continuous function which has this behaviour in a  limit ($\Delta \rar 0$), 
  \begin{equation}
\alpha (z) = 
\frac{1}{2} \left(
\alpha_{+} + \alpha_{-} 
\right)  + \frac{1}{2} \left(
\alpha_{+} - \alpha_{-} 
\right)  \tanh\left(\frac{z}{{\Delta}}  \right)  \  , \qquad (\Delta \rightarrow 0)  \  .
\end{equation}
}
\par \tl{Given the shape functions defined above (see equation (\ref{shape:function})) and defining $\delta = 1-{\chi}$ (see Figure \ref{eg:shape:spheroid}, \ref{eg:rod:spheroid}, \ref{eg:shape:rod}), the tangent unit vector to the swimmer surface is 
\yhy{\begin{align}
\hat{\bs{t}}(z)  \  =  \  
\begin{dcases}
\ \hat{\bs{e}}_z  \  , & \quad  z \in \left( - \chi, \chi \right)  \ ; \\
\quad \\
\ \frac{ - \epsilon  \  \mbox{sgn}(z) \Big(|z| -  {\chi} \Big) \ \hat{\bs{e}}_r  +  \sqrt{ \delta^4 - \delta^2 \ \Big( |z| -  {\chi} \Big)^2} \ \hat{\bs{e}}_z }{ \sqrt{ \delta^4 + \Big( \epsilon^2 - \delta^2 \Big) \Big( |z| -  {\chi} \Big)^2 }}   \ , &  \quad    z \in [-1,-{\chi}) \cup ( {\chi},1]  \ ,
\end{dcases}    \label{tangent:vector:generic}
\end{align}  }
where $\mbox{sgn}(x)$ is the sign function. By varying $\chi$ between 0 and 1 we can interpolate between a tapered rod and a blunt rod. \\
}
%
%
%
%
%
%
\subsubsection{\tl{Needle-like shape (Tapered ends)} }
\tl{We first consider the spheroidal swimmer geometry, i.e $\chi = 0$ (see Figure \ref{eg:shape:spheroid}), with shape function
\begin{equation}
S(z) = S(z; {\chi} = 0) = \sqrt{1-z^2}  \  ; \qquad    z \in [-1 , 1]  \  .
\end{equation}
 for which an exact solution exists\cite{popescu2010phoretic,nourhani2016geometrical}. For this shape, the unit tangent vector in equation (\ref{tangent:vector:generic}) simplifies to
\begin{equation}
\hat{\bs{t}}  \  =  \frac{\sqrt{1-z^2} \ \hat{\bs{e}}_z   -    \epsilon \, z \ \hat{\bs{e}}_r }{\sqrt{1 - z^2 + \epsilon^2 z^2}}   \quad ; \quad  \qquad  z \in [-1,1]    \  .  
\end{equation}
For a half-coated $(\alpha_{+}= - 1, \alpha_{-}=0)$ spheroid swimmer with uniform mobility ($\mu(z) = \mu_{+}=\mu_{-} = 1$), we obtain } 
\tl{ 
\begin{align}
U_{-1} & = - \frac{1}{2} \int_{-1}^{1} \left(  \hat{\bs{e}}_z \cdot \hat{\bs{t}} \right)    V_{-1}^{\text{slip}}(z) \ \mathrm{d}z  \  =    - \frac{1}{2} \int_{-1}^{1} \left(  \hat{\bs{e}}_z \cdot \hat{\bs{t}} \right)    \  \   \hat{\bs{t}}  \cdot \nabla \  C_I^{(-1)}(z) \ \mathrm{d}z  \   =  \  o(\epsilon)   \  ,  \\
U_0  &  =  - \frac{1}{2} \int_{-1}^{1} \left(  \hat{\bs{e}}_z \cdot \hat{\bs{t}} \right)    V_0^{\text{slip}}(z) \ \mathrm{d}z  \  =    - \frac{1}{2} \int_{-1}^{1} \left(  \hat{\bs{e}}_z \cdot \hat{\bs{t}} \right)  \  \    \hat{\bs{t}}  \cdot \nabla \  C_I^{(0)}(z) \ \mathrm{d}z  \    =  \frac{1}{2}  \  +  \   o(\epsilon)  \  ,
\end{align}  
where  $C_I^{(-1)}$ and $C_I^{(0)}$ are defined in equations (\ref{solute:expansion:1},\ref{solute:expansion:0}).}
\yhy{From eqn. (\ref{U:n:all}) all higher order terms vanish $U_n = 0$ ($n\geq 1$).} \\
Therefore, the speed is asymptotically,
\begin{equation}
U 
\ = \    \frac{1}{2} \ \epsilon  \  +  \  o\left( \epsilon^2 \, \log\left(\epsilon\right)\right) \   ,  \label{spheroid:asymptotic:speed}
\end{equation} 
which agrees with the asymptotic limit of the exact solution~\cite{nourhani2016geometrical} and the asymptotic result~\cite{schnitzer2015osmotic}.
%
%
%
\subsubsection{\yhy{Rod-like shape (Blunt ends)} }
\begin{figure}[h!]
\begin{center}
\subfigure[\quad high curvature end with $ 0 < \delta = 1-{\chi} < \epsilon \ll 1$.]{
\includegraphics[scale=.3]{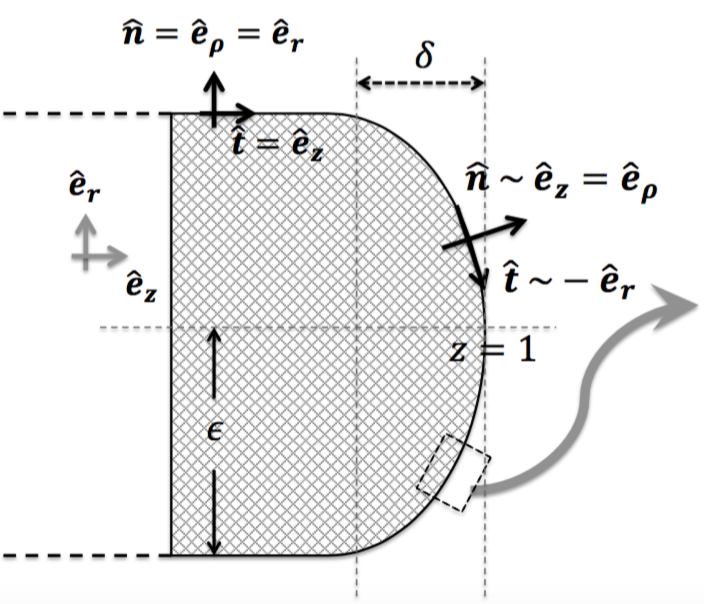}
}
\subfigure[\quad asymptotic regions.]{
\includegraphics[scale=.3]{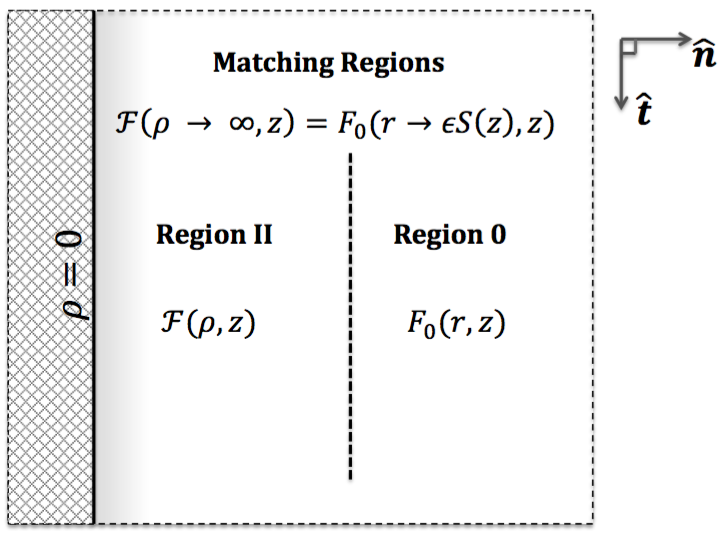}
}
\caption{Sketch of the swimmer \emph{blunt} end - showing the matching regions. $ \delta =  1-{\chi}$ is the width of the near end region where the shape function decays to zero. Note that the intermediate (region I) vanishes near the ends.} \label{blunt:end:sketch}
\end{center}
\end{figure}
\tl{We next consider a straight rod-like swimmer with blunt ends (see Figure \ref{blunt:end:sketch}). We define body shapes with \emph{blunt} ends as shapes having a small but finite end width $\delta= 1- \chi$, such that $0 < \delta  < \epsilon \ll 1$, where the body shape drops rapidly but continuously to zero (see Fig. \ref{blunt:end:sketch}). 
%
Thus, the projection of the tangent vector along the axis is }
 \begin{align}
\hat{\bs{e}}_z \cdot \hat{\bs{t}}  \  =  \  
\begin{dcases}
\   1 \  , & \quad  z \in \left( - \chi, \chi \right)  \ ; \\
\quad \\
\ o(\epsilon) \ , &  \quad    z \in [-1,-{\chi}) \cup ( {\chi},1]  \ ,
\end{dcases} 
\end{align}
\yhy{since $\textsf{maximum}\{ \delta/\epsilon \} = 1$.}
%
Hence, to the leading order, integrals involving the projection $\hat{\bs{e}}_z \cdot \hat{\bs{t}}$ are evaluated as 
\begin{equation}
\int_{-1}^{1} (\hat{\bs{e}}_z \cdot \hat{\bs{t}}) \quad \Big( \cdot  \Big) \quad dz \quad  = \quad  \int_{-{\chi}}^{{\chi}}   \quad \Big( \cdot  \Big) \quad  dz   + o\left( \epsilon \right), \qquad (0 < 1 - {\chi} < \epsilon \ll 1)  \  .  \label{blunt:ends:effect}
\end{equation}
\yhy{This is crucial in determining the propulsion speed since for these shapes $S(z=\pm {\chi})=1$ while $S(z=\pm 1) = 0$.}  \\
\par \tl{Considering as before a half-coated swimmer $(\alpha_{+}= - 1, \alpha_{-}=0 )$ with uniform mobility $\mu(z) = \mu_{+}=\mu_{-} = 1$, the first two terms in the asymptotic expansion are 
\begin{align}
U_{-1} & =  - \frac{1}{2} \int_{-1}^1  \left( \hat{\bs{e}}_z \cdot \hat{\bs{t}}\right)   V_{-1}^{\text{slip}} \  \mathrm{d}z \  =  - \frac{1}{2} \int_{-{\chi}}^{{\chi}} \frac{dC_I^{(-1)}}{dz} \mathrm{d}z   \   +  o(\epsilon)   \  =   \frac{1}{2} \ + \ o(\epsilon)  \  , \\
U_{0} & = - \frac{1}{2} \int_{-1}^{1}  \left( \hat{\bs{e}}_z \cdot \hat{\bs{t}} \right) \  V_0^{\text{slip}}(z;\epsilon) \ \mathrm{d}z -  2 \int_{-1}^1  q_0(z) \log{\left( \frac{2 \sqrt{1-z^2}}{S(z)} \right)}  \ \mathrm{d}z \nonumber  \\
& =  - \frac{1}{2} \int_{-{\chi}}^{{\chi}}  \frac{dC_I^{(0)}}{dz} \mathrm{d} z  -  2 \int_{-1}^1  q_0(z) \log{\left( \frac{2 \sqrt{1-z^2}}{S(z)} \right)}  \ \mathrm{d}z  \ + \ o(\epsilon) \  =  \   \left( \log{2} - \frac{1}{2} \right)  \   + \  o(\epsilon) \  .
\end{align}  }
Note that the integration limits are $z=\pm {\chi}$ rather than $z= \pm 1$ since the high curvature ends do not contribute to the propulsion velocity (see \yhy{equation \ref{blunt:ends:effect}}).
Hence, we obtain a speed 
\yhy{\begin{equation}
U  
 \  =  \  \frac{1}{2} \epsilon \left[  \log \left( \frac{4}{\epsilon}\right)  -1 \right] + \mathcal{O}\left( \frac{\epsilon}{\log(\frac{1}{\epsilon})}\right)  \  .  \label{rod:asymptotic:speed}
\end{equation} }
We note that with the same physico-chemical properties, this asymptotic speed is significantly \yhy{faster} than that of the \tl{spheroid shape (equation (\ref{spheroid:asymptotic:speed})). 
}\\
\subsubsection{\tl{Asymmetric shape (mixed ends)} }
\begin{figure}
\begin{center}
\includegraphics[scale=.35]{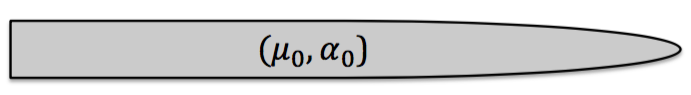}
\caption{Fore-aft asymmetric swimmer with uniform physico-chemical properties over its surface.}  \label{eg:rod:spheroid:uniform}
\end{center}
\end{figure}

\tl{It has recently been pointed out that geometric asymmetry with uniform coating can also lead to self-locomotion\cite{shklyaev2014non}.  With that in mind,
 we consider a slender swimmer with an asymmetric shape that lacks fore-aft symmetry (having one blunt end and one tapered end) but with uniform chemical activity over its surface $\mu(z)=\mu_{0} = 1, \ \alpha(z) = \alpha_{0} = - 1$ (see Figure \ref{eg:rod:spheroid:uniform})
\begin{align}
S(z) = 
\begin{dcases}
S(z; 0 < 1 - {\chi} < \epsilon ) ,  & - 1 \leq z \leq 0, \\
 S(z; {\chi} = 0 ) , & \quad 0 \leq z \leq 1.
 \end{dcases}
\end{align}
Hence we obtain the propulsion velocity at leading order in $\epsilon$, 
\begin{equation}
U = U_{-1}  \ \epsilon \ \log\left(  \frac{1}{\epsilon }\right)+ U_0 \ \epsilon + \cdots \  = \  \frac{1}{2} \epsilon \left[ \log \left( \frac{ \sqrt{2} }{\epsilon}\right)  - 1\right]    + \mathcal{O}\left( \frac{\epsilon}{\log(\frac{1}{\epsilon})}\right)   \ . 
\end{equation}
 }
\section{Discussion}
%
%
\tll{We have performed a detailed study of how the shape of slender diffusiophoretic self-propelled colloid (a phoretic swimmer) affects the speed at which the colloid 'swims'.  By slender here, we mean that the aspect ratio of average diameter to length is much less than 1, $\epsilon = a/l \ll 1$. In our calculations, the thickness of the phoretic slip layer, $L^*$ is taken as the smallest length in the problem since it is a molecular length-scale comparable to the interaction range of the reactant (product) molecules, while the diameter $a$ is a much larger colloidal length-scale, so we have the hierarchy of length-scales $l \gg a \gg L^*$.  Our approach is based on using simple axisymmetric 
shape functions which are able to interpolate between needle-like swimmers (spheroids) which have tapered ends and cylindrical rod-like swimmers with blunt ends. We note that cylindrical rod-like swimmers with blunt ends are easier to manufacture and hence have been the subject of many of the experimental studies to date. To obtain results in the limit $\epsilon \; \rar \; 0$, we have used the method of matched asymptotic expansions.} 

Our main result is the observation that the swimmer propulsion speed is strongly shape-dependent and in particular depends on whether the swimmer has pointed or blunt ends,  i.e has a needle-like or rod-like shape respectively. We show that swimmers with blunt ends (cylinders) have a higher propulsion speed than swimmers with tapered ends (needles) of the same aspect ratio, $\epsilon$. 

A simple argument explaining the differences in swimming efficiency can be obtained by considering the competition of effects that lead to self-propulsion.
Self-propulsion arises from the balance of the hydrodynamic stresses due to the slip velocity generated by the concentration gradients of reactants (products) and the associated hydrodynamic drag due to fluid velocity gradients near the surface of the swimmer. The constraint of force (torque) balance on the swimmer then selects  a unique propulsion velocity. For a given shape, the higher the concentration gradient developed, the higher the propulsion velocity. The concentration gradients from the reaction occur on a length-scale which is set by the length of the swimmer and is mostly independent of the detailed shape of the swimmer, however the slip and drag they generate depend on the local surface area of the swimmer and are sensitive to the detailed shape of the swimmer.  For a spheroid, the hydrodynamic stresses leading to drag are comparable to the stresses due to slip along the whole surface of the colloid, while for the blunt cylinder, the effects of phoretic slip dominate the drag at the highly curved ends~\cite{nourhani2016geometrical}. 
At a fixed aspect ratio, $\epsilon$, the surface area of the spheroid, $A_{sph}$ is larger than the surface area of the blunt cylinder, $A_{cyl}$. Hence the balance of stresses is skewed towards a higher propulsion velocity for blunt cylinders.

Our results clearly imply that care must be taken in designing theoretical models for self-phoretic slender swimmers and that slender self-phoretic cylindrical rods cannot be  accurately modelled by slender self-phoretic spheroids. We note that there is a limit when such details of the swimmer shape do not matter - when the slip layer is larger than the diameter of the rods. However this limit, as discussed above, is unphysical.

\begin{acknowledgments}
This work was supported by EPSRC grant EP/G026440/1 (TBL, YI), and HFSP grant RGP0061/2013 (RG). YI acknowledges the support of University of Bristol. TBL acknowledges support of BrisSynBio, a BBSRC/EPSRC Advanced Synthetic Biology
 Research Centre (grant number BB/L01386X/1).
\end{acknowledgments}

\appendix

\section{Asymptotic analysis}
\label{app:A}
\subsection{Region 0: Outer region} \label{app:reg0}
\tl{On length scales far from the swimmer  (i.e $ r \gg \epsilon $), the swimmer is a line of vanishing thickness of length $2$. Therefore, we use the standard slender body approximation~\cite{Batchelor70Slender,cox1970motion,keller1976slender,geer1976stokes}, for the concentration and velocity fields using distributions of line singularities along the swimmer centreline. 
%
In this region, we only need the leading,  i.e. $\mathcal{O}(1) \equiv \mathcal{O}(\epsilon^0, \, \lambda^0)$ concentration and velocity fields for our analysis: 
\begin{align}
c(r,z) \  & = \ C_0(r,z)  \   , \\
\bs{v}(r,z) \ & = \ \bs{V}_0(r,z) \  .
\end{align}
Since, $ r \gg \epsilon $, the short-range interaction potential of the reactive species with the swimmer surface vanishes in this region
\begin{equation}
\psi(r - \epsilon S(z), z) \  = \  \Psi_0(r- \epsilon S(z), z) = 0 \ .
\end{equation}
}
\par 
\tl{We approximate the concentration by one generated by a line distribution of point sources~\cite{schnitzer2015osmotic}, along the swimmer centerline $(r,z) = (0,\xi)$, where $-1 \le \xi \le 1$.}
\begin{equation}
C_0(r,z) = C_{\infty} +  \int_{-1}^1 \frac{j({\xi})}{\sqrt{r^2 + (z - {\xi})^2}} \ d{\xi} , \label{solute:field:line:sing}
\end{equation}
where $j(z)$ is the \tl{source line density along the body centerline. It is to be determined by matching $C_0 (r,z)$ as $r \; \rar \;  0$ with the intermediate} (region I) expansion of concentration field (see Fig.  \ref{fig:slender:asymp:regions}). \tl{We note that since $\Psi_0(r- \epsilon S(z),\, z) = 0$ in this region, the concentration $C_0 (r,z)$,   is a solution of the steady state diffusion equation (\ref{gov:mass:transfer}). } \\

\tl{While the integral in equation (\ref{solute:field:line:sing}) is singular as $r \rightarrow 0$ and $|z| \leq 1$, 
there are a number of ways to isolate the singularity and approximate the integral analytically as the observation point approaches the swimmer surface. A particularly simple method is subtracting the singular part of the integral which gives the approximate expression~\cite{keller1976slender} }
\begin{align}
C_0(r,z) & =  C_{\infty} 
+ 2 j(z) \log \left( \frac{2}{r} \right) + \int_{-1}^1 \frac{j({\xi}) - j(z)}{|z-{\xi}|} \, \mathrm{d}{\xi}  + 2j(z) \log \left( \sqrt{1 - z^2} \right)  + \mathcal{O}\left( r^2\right)  \  . \label{solute:asymp:first}
\end{align}
\tll{For convenience, we integrate by parts the non-local  term in equation (\ref{solute:asymp:first}) to obtain an equivalent expression~\cite{tuck1964some}
\begin{equation}
C_0(r,z) = C_{\infty} + 2 j(z) \log\left( \frac{2}{r} \right)  - \int_{-1}^{1} \ \frac{\mathrm{d}}{\mathrm{d}\xi}j(\xi)  \    \mbox{sgn}(z-{\xi})  \,  \log|z-{\xi}|  \,  \mathrm{d}{\xi} + \mathcal{O}\left( r^2 \log(r) \right)  \  .  \label{solute:approx:region:0}
\end{equation} 
Note that we assumed $j(\pm 1) = 0 $ in obtaining the above expression, which we confirm to be  the case \emph{a posteriori} for our generic shape function defined in equation (\ref{shape:function}). }  \\
%
%
\par 
%
\tl{We approximate the fluid velocity field by one generated by a line distribution of point force singularities along the swimmer centerline~\cite{Batchelor70Slender} }
\begin{align}
\bs{V}_0 (r,z) & = - U \hat{\bs{e}}_z + \int_{-1}^1 \mathrm{d}{\xi} \ \left( \frac{ \hat{\bs{e}}_z }{\left|\bs{x} - \bs{x}' \right|}  + \frac{(z - {\xi})(\bs{x} - \bs{x}') }{\left|\bs{x} - \bs{x}' \right|^3} \right) q({\xi};\epsilon) \ ,  \label{flow:field:bulk}
\end{align}
where $\bs{x} - \bs{x}' = r \ \bs{\hat{e}}_r + (z - {\xi}) \ \bs{\hat{e}}_z$ and $q(z;\epsilon)$ is a \tl{generic force line density} along the body's centerline. 
$q(z;\epsilon)$ will  be determined by matching $\bs{V}_0(r,z)$ as $r \; \rar \;  0$ with the intermediate (region I) velocity field (see Fig.  \ref{fig:slender:asymp:regions}).
\tll{This velocity field is a solution of the Stokes equations (\ref{gov:continuity},\ref{gov:stokes}) since the short-range interaction potential vanishes in this region. There is an associated hydrostatic pressure, 
\beq
p(r,z) - p_\infty  = P_0 (r,z) = 2 \int_{-1}^1  \mathrm{d}\xi  \ \left(  \frac{ z-\xi }{\left | \bs{x}  - \bs{x}'  \right|^3} \right) \, q(\xi ; \epsilon ) \quad 
\nonumber
\eeq
} 

\tl{As for the concentration field discussed above (see equation (\ref{solute:field:line:sing})), the flow field  $\bs{V}_0 (r,z) $ is also singular as  $r \rightarrow 0$ and $ |z| \leq 1 $. Isolating the singular part in a similar way, the axial component of the fluid velocity has the asymptotic form\cite{tuck1964some}
\begin{equation}
V_{0,z}(r,z) = - U  + 4 q(z;\epsilon) \log\left( \frac{2 \sqrt{1-z^2}}{r} \right) - 2q(z;\epsilon) - 2 \int_{-1}^{1} \frac{q({\xi};\epsilon) - q(z;\epsilon)}{| z - {\xi} |} \, \mathrm{d}{\xi} + \mathcal{O}\left( r^2 \right).  \label{flow:approx:region:0}
\end{equation}   }
%
\par \tl{In the next section, we will solve for the concentration and velocity fields in the intermediate region I (see Fig. \ref{fig:slender:asymp:regions}). } \\

%
%
%
\subsection{Region I: \yhy{Intermediate region}}  \label{app:regI}
\tll{At intermediate length-scales, for which the radial distance is larger than the range of the short-ranged interaction but much less than the length of the swimmer, the swimmer is locally approximated by an infinite cylinder of thickness $2 \epsilon$. Here, we re-scale the radial coordinate with the swimmer slenderness ratio ($\epsilon = a/l$);}
\begin{equation}
 R \ = \  \frac{r}{\epsilon} \quad .
\end{equation}
\tl{In this region, the leading $\mathcal{O}(1) \equiv \mathcal{O}(\epsilon^0, \, \lambda^0)$ fields are 
\begin{align}
c(r,z) \ = \ C_I(R, z)  \  , \\
\bs{v}(r,z) \ = \ \bs{V}_I(R, z)   \  . 
\end{align}
and the short-range interaction between the molecules and the swimmer surface also  vanishes in this region (we have set $\lambda=0$) 
\begin{equation}
\psi(r - \epsilon S(z), z) \  = \  \Psi_I(R - S(z), z) = 0 \ .
\end{equation}
This is of course because we have assumed $\lambda \ll \epsilon$ (i.e the swimmer cross-sectional radius is much larger than the interaction range of the potential). } \\
\par \tl{Using the definition of the unit vectors, $\hat{\bs{t}},\hat{\bs{n}}$ in equations (\ref{unit:vector:normal}) and (\ref{unit:vector:tangent}), the local gradients (away from the ends) in the re-scaled coordinates~\cite{van1975perturbation}} are 
\begin{align}
\hat{\bs{n}}\cdot \nabla & \ =  \  \frac{1}{\epsilon} \frac{\partial}{\partial R} + {\cal O} \left(\epsilon\right)  \  ,  \label{normal:gradient:approximation} \\
\hat{\bs{t}} \cdot \nabla & \  =  \  \frac{\partial}{\partial z} + S_z(z) \frac{\partial}{\partial R} + {\cal O} \left(\epsilon\right)  \  , \label{tangent:gradient:approximation}
\end{align}
\\
%
%
\par \tl{\emph{(Concentration)}: The current $\bfJ= J_{I,z} \hat{\bs{e}}_z +J_{I,R}  \hat{\bs{e}}_r $ 
where $\hat{\bs{t}} \simeq\hat{\bs{e}}_z$ can be written in components (see equation (\ref{gov:mass:transfer})), 
\begin{align}
 J_{I,R} (R,z) =  -  \frac{\partial C_I(R,z)}{\partial R}\quad ; \qquad J_{I,z}(R,z) =  -  \frac{\partial C_I(R,z)}{\partial z}  \  .
\end{align}
Since the interaction vanishes, the equation of motion for the reactant (product) concentration is the steady state diffusion equation (\ref{gov:mass:transfer}), to  leading order has only radial contributions : 
\begin{align}
 0 & =  \ \epsilon^{-1}  \ \frac{1}{R}  \ \frac{\partial }{\partial R} \left( R \ \frac{\partial C_I}{\partial R} \right) \;  ; 
 \quad   \mbox{with boundary condition} \quad - \left. \frac{\partial C_I}{\partial R} \right|_{R = S(z)} \  =  \  J_{I,R}(S(z),z)  \  ,  \label{region1:diffusion}
\end{align}
where in the boundary condition, the (unknown) radial flux $J_{I,R}(S(z),z)$ is  
determined by matching $C_I(R,z), J_{I,R}(R,z)$ with the inner interaction layer (region II) on the swimmer surface. The other boundary condition is obtained from matching with the outer region 0.  } 
%
\tl{Therefore, solving equation (\ref{region1:diffusion}) and imposing the boundary condition, 
}
\begin{equation}
C_I(R,z;\epsilon) = A(z;\epsilon) +   J_{I,R}(S(z),z) S(z) \log \left( \frac{1}{R}  \right) , \label{solute:field:intermediate}
\end{equation}
where $A(z;\epsilon)$ is an 
unknown function of $z$  resulting from integrating over $R$, must be determined by matching with the outer (region 0) \tll{concentration, $C_0(r,z)$ given in equation (\ref{solute:approx:region:0}). 
Using this result, equation (\ref{summary:solute:expansion:1})  and the second term in equation (\ref{summary:solute:expansion:0})  are determined by $J_{I,R}$, while  the first and third terms in equation (\ref{summary:solute:expansion:0}) are determined by $A(z;\epsilon)$. 
We note that while $A$ is determined by the properties of the concentration as $r,z \; \rar \;  \infty$, $J_{I,R}$ depends on the concentration near the reactive colloidal surface.
}\\
\par  \tl{\emph{(Fluid velocity)}: The Stokes equations (\ref{gov:continuity},\ref{gov:stokes}) with $\Psi_I = 0$, at leading order involve only the axial component of the velocity field , $\bs{V} (R,z) = V_{I,z}(R,z) \, \hat{\bs{e}}_z +V_{I,R} (R,z) \,  \hat{\bs{e}}_r $ where $\hat{\bs{t}} \simeq\hat{\bs{e}}_z, \, \hat{\bs{n}} \simeq\hat{\bs{e}}_r$ :
\begin{align}
0 & = \frac{1}{R} \ \frac{\partial }{\partial R} \left( R \ \frac{\partial V_{I,z}}{\partial R} \right)  \  ;  \quad  \mbox{with slip boundary condition} \quad 
\left. V_{I,z} \right|_{R=S(z)} \  =  \   \hat{\bs{e}}_z \cdot \bs{V}^{\text{slip}}(z;\epsilon)  \  .
\label{Stokes:interm} \end{align}}\tll{
The hydrostatic pressure difference, $p(R,z) - p_\infty = P_I(R,z)$ vanishes at this order.
The slip velocity, $\bs{V}^{\text{slip}}(z;\epsilon) $ is obtained by matching with the inner interaction layer (region II). Solving equation (\ref{Stokes:interm}) gives,}
\begin{equation}
V_{I,z} (R,z;\epsilon)  = \hat{\bs{e}}_z \cdot \bs{V}^{\text{slip}}(z;\epsilon)  + \mathcal{W}(z;\epsilon) \ \log{\left( \frac{R}{S(z)} \right)}   \  .   \label{flow:field:intermediate}
\end{equation}
\tl{The unknown function, $\mathcal{W}(z;\epsilon)$ resulting from integrating over $R$ will be found by matching with the  fluid velocity in region 0, far from the swimmer surface.}
  \\
%
%
\subsection{Region II: Inner interaction (boundary) layer}\label{app:regII}

\tl{For a swimmer with cross-sectional radius much larger than the interaction range of the potential, $\psi$ between reactant (product) molecules and the swimmer surface  ($L^* \ll a$),  the swimmer surface is locally flat, and the length scale of the variation of fields is set by the small parameter $\lambda = L^*/l$ with $2l$ being the length of the swimmer. Now, provided we are away from the  ends of the rod where the curvature is high, the normal unit vector to the surface is $\bs{\hat{n}} \sim \bs{\hat{e}}_r$. Hence, we re-scale the radial coordinate as
\begin{equation}
\rho = \frac{r - \epsilon \ S(z)}{\lambda}; \quad \mbox{where} \quad  \lambda \ = \ \frac{L^*}{l}, \quad \epsilon \  = \  \frac{a}{l}, \quad \lambda \ll \epsilon  \  .
\label{eq:coords_inner}\end{equation}
%
%
In this region,  we must perform an expansion in both of the small parameters $\lambda$ and $\epsilon$ (noting our assumption $L^* \ll a$ or equivalently $\lambda \ll \epsilon$).
We therefore express the fields (concentration, fluid velocity and pressure) w.l.g. as 
\begin{align}
c(r,z) & = \mathcal{C}(\rho,z)  = \mathcal{C}^{(0,0)}(\rho,z)+ \epsilon \ \mathcal{C}^{(0,1)}(\rho,z) + \lambda \ \mathcal{C}^{(1,0)}(\rho,z) + \lambda \epsilon \ \mathcal{C}^{(1,1)}(\rho,z) +  \cdots \label{eq:c_exp_dble} \\
\bs{v}(r,z) & = \bs{\mathcal{V}}(r,z)  = \bs{\mathcal{V}}^{(0,0)}(\rho,z) + \epsilon \ \bs{\mathcal{V}}^{(0,1)}(\rho,z) + \lambda \ \bs{\mathcal{V}}^{(1,0)}(\rho,z) + \lambda \epsilon \ \bs{\mathcal{V}}^{(1,1)}(\rho,z) + \cdots \label{eq:v_exp_dble}  \\
p(r,z) & = \mathcal{P}(r,z)  = \lambda^{-2} \ \mathcal{P}^{(-2,0)}(\rho,z) + \lambda^{-1} \ \mathcal{P}^{(-1,0)}(\rho,z)  + \epsilon \lambda^{-2} \ \mathcal{P}^{(-2,1)}(\rho,z) + \epsilon\lambda^{-1} \ \mathcal{P}^{(-1,1)}(\rho,z) + \cdots \label{eq:p_exp_dble} 
\end{align}
The leading terms for the concentration are of order ${\cal O} (\epsilon^0)$, while the leading order terms of velocity and pressure are of  ${\cal O} (\epsilon)$. Therefore for an asymptotic analysis, we can simplify the expansions above. We can set $\mathcal{C}^{(0,1)}(\rho,z) =  \mathcal{C}^{(1,1)}(\rho,z)=0$ and rename $\mathcal{C}^{(0,0)}(\rho,z) \equiv \mathcal{C}^{(0)}(\rho,z)$ and $\mathcal{C}^{(1,0)}(\rho,z) \equiv \mathcal{C}^{(1)}(\rho,z)$. Similarly we can set $\bs{\mathcal{V}}^{(0,0)}(\rho,z) = \bs{\mathcal{V}}^{(1,0)}(\rho,z) =0$, $\mathcal{P}^{(-2,0)}(\rho,z) =\mathcal{P}^{(-1,0)}(\rho,z) =0$ and rename $\bs{\mathcal{V}}^{(0,1)}(\rho,z) \equiv \bs{\mathcal{V}}^{(0)}(\rho,z)$,  $\bs{\mathcal{V}}^{(1,1)}(\rho,z)  \equiv \bs{\mathcal{V}}^{(1)}(\rho,z) $, $\mathcal{P}^{(-2,1)}(\rho,z) \equiv \mathcal{P}^{(-2)}(\rho,z) $ and $\mathcal{P}^{(-1,1)}(\rho,z) \equiv \mathcal{P}^{(-1)}(\rho,z) $.  }

\par \tl{\emph{(Concentration)}:
We can therefore write out the concentration and interaction potential as expansions in in the small parameter $\lambda$ :
\begin{align}
c(r,z) & = \mathcal{C}(\rho,z)  = \mathcal{C}^{(0)}(\rho,z) + \lambda \ \mathcal{C}^{(1)}(\rho,z) + \cdots  \\
\psi(r-\epsilon S(z),z) & = \Psi(\rho,z) = \Psi^{(0)}(\rho,z) + \lambda \ \Psi^{(1)}(\rho,z) + \cdots  
\label{eq:pot_expand}\end{align}
with associated flux written in components (see equation (\ref{gov:mass:transfer})),  $\bfJ (r,z) = \bs{\mathcal{J}}(\rho, z)  = \mathcal{J}_{z}\,  \hat{\bs{e}}_z + \mathcal{J}_{\rho}  \, \hat{\bs{e}}_r $. Away from the ends $\hat{\bs{t}} \simeq \hat{\bs{e}}_z , \, \hat{\bs{n}} \simeq\hat{\bs{e}}_r$ and 
the leading contribution to the flux is radial, 
\begin{equation}
 {\mathcal{J}}_\rho(\rho, z) \  = \  \lambda^{-1} \   {\mathcal{J}}_\rho^{(-1)}(\rho, z)   \   +  \  {\mathcal{J}}_\rho^{(0)}(\rho, z)  \  + \  \mathcal{O}(\lambda)  \  .
\end{equation}
where  
\begin{align}
\mathcal{J}_{\rho}^{(-1)} & = - \ \epsilon \left[ \frac{\partial \mathcal{C}^{(0)}}{\partial {\rho}} + \mathcal{C}^{(0)} \frac{\partial \Psi^{(0)}}{\partial {\rho}} \right] \quad  ;  \qquad
\mathcal{J}_{\rho}^{(0)}  = - \ \epsilon \left[ \frac{\partial \mathcal{C}^{(1)}}{\partial {\rho}} + \mathcal{C}^{(0)} \frac{\partial \Psi^{(1)}}{\partial {\rho}} + \mathcal{C}^{(1)} \frac{\partial \Psi^{(0)}}{\partial {\rho}}\right]  \quad  \quad   \  .
\label{eq:flux_inner}\end{align} 
Hence, equation (\ref{gov:mass:transfer}) becomes 
\begin{equation}
\lambda^{-2} \ \frac{\partial \mathcal{J}_\rho^{(-1)}}{\partial {\rho}}  + \lambda^{-1} \left[ \frac{  \mathcal{J}_{\rho}^{(-1)}  }{\left( \lambda {\rho} + \epsilon S(z) \right) } + \frac{\partial \mathcal{J}_{\rho}^{(0)}}{\partial {\rho}} \right]  + \mathcal{O}\left( 1 \right) = 0   \  ,
\label{eq:inner_mass_transfer}\end{equation}
with the corresponding boundary condition 
\begin{equation}
 \Big[ \lambda^{-1} \ {\mathcal{J}}_\rho^{(-1)}  +  \ {\mathcal{J}}_\rho^{(0)} + \mathcal{O}(\lambda)  \Big]_{{\rho} = 0}  \   =  \   \alpha(z)   \  .
\end{equation}
Therefore at leading order in the flux, equation (\ref{gov:mass:transfer}) becomes 
\begin{align}
\frac{\partial \mathcal{J}_{\rho}^{(-1)}}{ \partial {\rho}} & = 0  \quad ; \quad \mbox{with boundary condition} \quad 
\left. \mathcal{J}_{\rho}^{(-1)} \right|_{{\rho} =0} \  =  \  0   \  .
\label{eq:lead_flux_inner} \end{align}
Integrating equation (\ref{eq:lead_flux_inner}) once gives $\mathcal{J}_{\rho}^{(-1)} = 0$. Integrating again, using the definition of the flux, equation (\ref{eq:pot_expand})  and equation (\ref{eq:flux_inner}), we obtain an expression for the concentration in the inner region~\cite{anderson1982motion}
\begin{equation}
 \mathcal{C}^{(0)}({\rho},z;\epsilon) = C_s(z;\epsilon) \exp{\left( - \Psi^{(0)}({\rho},z) \right)} \quad ; \qquad C_s(z;\epsilon) = \mathcal{C}^{(0)}(\rho \rightarrow \infty,z) \  . \label{inner:solute:field}
\end{equation}
The as yet undetermined integrating constant, $C_s(z;\epsilon)$ will be found by matching with the 
concentration, $C_I(R,z)$ in the intermediate region.  } \\
\par \tl{To complete the matching with the intermediate region, we require the next to leading order terms in the flux, i.e.
$\mathcal{O}(\lambda^0)$, 
which is obtained by solving 
equation  (\ref{eq:inner_mass_transfer}) at $\mathcal{O}(\lambda^{-1})$ : 
\begin{align}
\frac{\partial \mathcal{J}_{\rho}^{(0)}}{\partial \rho} & = 0 \quad ; \quad \mbox{with boundary condition} \quad 
\left. \mathcal{J}_{\rho}^{(0)} \right|_{\rho = 0} \  = \  \alpha(z)  \  .
\end{align}
It follows then that 
\begin{equation}
\mathcal{J}_{\rho}^{(0)}(\rho=0,z) \  =  \  \lim_{\lambda \rightarrow 0, \ \rho \rightarrow \infty} \mathcal{J}_{\rho}^{(0)}(\rho,z)  \   =  \  \alpha(z)  \  , \label{flux:regionII}
\end{equation}
which will provide the required boundary condition for matching with the flux in the  intermediate (region I). } \\
%

\par \emph{(Fluid velocity)}: \tl{Returning to equations (\ref{gov:continuity}, \ref{gov:stokes}), we use equations (\ref{eq:c_exp_dble},\ref{eq:v_exp_dble},\ref{eq:p_exp_dble}), to express the velocity and pressure as expansions in the small parameter $\lambda$ (note that they both vanish when $\epsilon=0$),
\begin{align}
\bs{v}(r,z) & = \bs{\mathcal{V}}(r,z)  = \epsilon \, \bs{\mathcal{V}}^{(0)}(\rho,z) + \epsilon \,  \lambda \ \bs{\mathcal{V}}^{(1)}(\rho,z) + \cdots \\
p(r,z) & = \mathcal{P}(r,z)  = \epsilon \,  \lambda^{-2} \ \mathcal{P}^{(-2)}(\rho,z) + \epsilon \,  \lambda^{-1} \ \mathcal{P}^{(-1)}(\rho,z) + \cdots \ .
\end{align}
It is convenient to express equations (\ref{gov:stokes}) in cylindrical coordinates for axisymmetric velocity fields, 
\beq
\frac1r {\partial \over \partial r} \left( r {\partial \over \partial r}  \bs{\mathcal{V}} \right) 
 +  \left( {\partial^2 \over \partial z^2} \bs{\mathcal{V}} \right)
-  \left(  \nabla {\mathcal{P}} +  \epsilon \lambda^{-2} {\mathcal{C}} \nabla \Psi  \right)   = \bf 0
\eeq
or alternatively
\beq
\frac1r {\partial \over \partial r} \left( r {\partial \over \partial r}  \bs{\mathcal{V}} \right) 
 +  \left( {\partial^2 \over \partial z^2} \bs{\mathcal{V}} \right)
-  \hat{\bs{e}}_z \left(  {\partial {\mathcal{P}} \over \partial z} +  \epsilon \lambda^{-2} {\mathcal{C}} {\partial \Psi \over \partial z} \right)-  \hat{\bs{e}}_r   \left( {\partial {\mathcal{P}} \over \partial r} +  \epsilon \lambda^{-2} {\mathcal{C}} {\partial \Psi \over \partial r} \right)    = \bf 0
\eeq
%
%
Using equations (\ref{eq:coords_inner},\ref{eq:c_exp_dble},\ref{eq:v_exp_dble},\ref{eq:p_exp_dble}) and projecting the equations onto the local inner coordinates $( \hat{\bs{t}}(z), \hat{\bs{n}}(z))$ which are functions of $z$ only, $\bs{\mathcal{V}}^{(0)}= {\mathcal{V}}_n^{(0)}  \hat{\bs{n}} + {\mathcal{V}}_t^{(0)}  \hat{\bs{t}}$ (see Fig. \ref{blunt:end:sketch}) :
\begin{align}
 \hat{\bs{n}}: \qquad \qquad  0 &   \ = \epsilon \, \left ( \ - \lambda^{-3}  \    \hat{\bs{n}} \cdot  \hat{\bs{e}}_r \left[ \frac{\partial \mathcal{P}^{(-2)}}{\partial {\rho}} +  \ \mathcal{C}^{(0)} \frac{\partial \Psi^{(0)}}{\partial {\rho}} \right] + \lambda^{-2} \frac{\partial^2 {\mathcal{V}}^{(0)}_{n}}{\partial {\rho}^2} - \lambda^{-2}  \    \hat{\bs{n}} \cdot  \hat{\bs{e}}_z \left[ \frac{\partial \mathcal{P}^{(-2)}}{\partial {z}} +  \ \mathcal{C}^{(0)} \frac{\partial \Psi^{(0)}}{\partial {z}} \right]  \right . \\ 
&  \left.  \qquad \quad  + \;  \frac{\lambda^{-1}}{\left( \lambda {\rho} + \epsilon S(z) \right)} \frac{\partial {\mathcal{V}}^{(0)}_{n}}{\partial {\rho}} + \mathcal{O}(\lambda^0)   \ \right) ,  \label{gov:stokes:expansion:normal} \\
\hat{\bs{t}}: \qquad \qquad  0 &  \  =  \epsilon \, \left ( \  \lambda^{-2}  \   \left[ \frac{\partial^2 }{\partial {\rho}^2}  \hat{\bs{t}} \cdot \bs{\mathcal{V}}^{(0)} - \hat{\bs{t}} \cdot \nabla \mathcal{P}^{(-2)} -  \ \mathcal{C}^{(0)} \hat{\bs{t}} \cdot \nabla \Psi^{(0)} \right] +  \frac{\lambda^{-1}}{\left( \lambda {\rho} + \epsilon S(z) \right)} \frac{\partial }{ \partial {\rho}} {\mathcal{V}}_{t}  + \mathcal{O}(\lambda^0)  \ \right) ,  \label{gov:stokes:expansion:tangential}
\end{align}
while equation (\ref{gov:continuity}) becomes at leading order in $\lambda$
\begin{equation}
\epsilon \, \left ( \  \lambda^{-1} \ \frac{\partial {\mathcal{V}}_{n}^{(0)}}{\partial \rho}  +  \mathcal{O}(\lambda^0)  \  \right) =  \  0  \ .  \label{incompressiblity:2}
\end{equation}
The leading order, $\mathcal{O}(\lambda^{-3})$  terms in equation (\ref{gov:stokes:expansion:normal})  imply that the stresses induced by the interaction of the reactant (product) molecules with the swimmer surface are balanced by a local radial pressure gradient, 
\begin{equation}
\frac{\partial \mathcal{P}^{(-2)}}{\partial {\rho}} +  \ \mathcal{C}^{(0)} \frac{\partial \Psi^{(0)}}{\partial {\rho}} = 0  \  . \label{radial:momentum:balance}
\end{equation}
Solving equation (\ref{radial:momentum:balance}), and using the already calculated $\mathcal{C}^{(0)}$ given in equation (\ref{inner:solute:field}), we obtain the hydrostatic pressure profile \cite{anderson01,Yariv2011}
\begin{equation}
\mathcal{P}^{(-2)}({\rho},z;\epsilon) =  \ C_I(S(z),z;\epsilon) \left[ e^{- \Psi^{(0)}({\rho},z) } - 1 \right] \  .   \label{inner:pressure} 
\end{equation}
To obtain it we have matched with the pressure in the intermediate (region I) which requires that $\ds \lim_{\rho \rar \infty}\mathcal{P}^{(-2)}(\rho, z) = 0$  and $\ds \lim_{\rho \rar \infty} \Psi^{(0)} ({\rho},z) = 0$. 
%
In the tangential direction, at leading order, i.e. $\mathcal{O}(\lambda^{-2})$, in equation (\ref{gov:stokes:expansion:tangential}),  the  viscous stresses balance the pressure gradient (due to the concentration gradient) and the tangential stresses  induced by the interaction of the molecules of reactant (product)  with the swimmer surface
\begin{equation}
 \frac{\partial^2 }{\partial {\rho}^2} {\mathcal{V}}_{t} - \hat{\bs{t}} \cdot \nabla \mathcal{P}^{(-2)} -  \ \mathcal{C}^{(0)} \hat{\bs{t}} \cdot \nabla \Psi^{(0)}  = 0  \ . \label{tangential:momentum:balance}
\end{equation}
}
\tll{In addition, at the next to leading order, $\mathcal{O}(\lambda^{-2})$ in equation (\ref{gov:stokes:expansion:normal})
\begin{equation}
\frac{\partial^2 {\mathcal{V}}_{n}^{(0)} }{\partial \rho^2}  \  =  \   0  \ ,  \label{radial:momentum:balance:2}
\end{equation}
since from equation (\ref{unit:vector:normal}) , $\hat{\bs{n}} \cdot  \hat{\bs{e}}_z = O(\epsilon)$.
Note that $\bs{\hat{t}} \sim \bs{\hat{e}}_z$ away from the ends and $\bs{\hat{t}} \sim \bs{\hat{e}}_{\rho}$ at the ends (see Fig. \ref{blunt:end:sketch}).
Now, using  equation (\ref{inner:pressure}) and  equation (\ref{inner:solute:field}), we can integrate equation (\ref{tangential:momentum:balance}) twice to obtain  the tangential velocity and hence the limiting value 
\begin{equation}
\lim_{\lambda \rightarrow 0; \ \rho \rightarrow \infty} {\mathcal{V}}_{t}  \   =  \   \epsilon \ \mu(z) \ \hat{\bs{t}} \cdot \nabla C_I(z;\epsilon)  \  \  . 
\end{equation}
In addition, from equations (\ref{radial:momentum:balance:2},\ref{incompressiblity:2}), the normal component of the velocity vanishes, ${\mathcal{V}}_{n}^{(0)} = 0$. Thus,  we obtain the slip velocity~\cite{anderson01}
\begin{equation}
  \bs {V}^{\text{slip}}(z;\epsilon) \  =  \     \epsilon \ \mu(z) \  \hat{\bs{t}} \cdot \nabla C_I(z;\epsilon)  \  \hat{\bs{t}} \    \  ,  \label{slip:velocity}
\end{equation}
where we define the diffusiophoretic mobility~\cite{anderson01} $\mu(z) = \int_0^\infty \ {\rho} \left[ 1  - e^{- \Psi^{(0)}({\rho},z) } \right] \mathrm{d}{\rho} .$  } \\
\tl{In the next section, we match the solutions from the three regions to find the unknown functions $J_{I,R}(R,z), A(z;\epsilon),j(z)$, $\mathcal{W}(z;\epsilon)$ and $q(z)$ arising from our integrations.}\\
%
%

\subsection{Asymptotic matching}
\label{app:asymp}
\par \emph{(Concentration)}: First, we match the \tl{$\mathcal{O}(\epsilon^0, \lambda^0)$ fluxes between regions I and II;}
\begin{equation}
 \mbox{(II)} \quad \lim_{\yhy{\lambda \rightarrow0 , } \, \rho \rightarrow \infty}   \   \mathcal{J}_{\rho}^{(0)}(\rho,z)   \ \  =  \  \  \alpha(z)  \   \    =   \   \    \lim_{R \rightarrow S(z)}  \  J_{I,R}(R,z)   \  \quad \mbox{(I)} \;   ,
\end{equation}
\tll{using equation (\ref{flux:regionII}).}
 \tl{Next, we match the  $\mathcal{O}(\epsilon^0, \lambda^0)$ concentrations between intermediate (region I) with the outer (region 0).} 
\tll{Using equations (\ref{solute:approx:region:0} and \ref{solute:field:intermediate}) }
\begin{equation}
 \mbox{(I)} \quad  C_I(R,z;\epsilon)  \  \   =  \   \    A(z;\epsilon) + \alpha(z) S(z) \log \left( \frac{1}{R} \right)   \   \  =   \    \   \lim_{r  \rightarrow  \epsilon R \ll 1} C_0(r,z) \quad  \qquad \mbox{(0)} \quad . \label{solute:matching:cyl:bulk}
\end{equation}
\tll{we obtain  the source line density $j(z)$, introduced in equation (\ref{solute:field:line:sing})  and the unknown function $A(z;\epsilon)$ introduced in equation (\ref{solute:field:intermediate}) } :
\begin{equation}
j(z) = \frac{1}{2} \alpha(z)S(z)  \  ; \qquad A(z;\epsilon)   =   C_{\infty} + 
\alpha(z) S(z) \log \left( \frac{2}{\epsilon} \right)  -  \frac{1}{2} \int_{-1}^1 \frac{\mathrm{d}}{\mathrm{d} \xi} \Big[\alpha({\xi})S({\xi})\Big] \ \mbox{sgn}\left( z - {\xi} \right) \log|z-{\xi}|  \ d{\xi}  \  .
\end{equation}
Hence we obtain the leading order concentration in region I 
\begin{equation}
C_I(R,z;\epsilon) =  C_I^{(-1)} (z) \log \left( \frac{1}{\epsilon}\right)  + C_I^{(0)}(z)  + \alpha(z) S(z) \log \left( \frac{S(z)}{R} \right)  \ ,  \label{solute:field:intermediate:matching}
\end{equation}
where \tl{we have defined}
\begin{align}
C_I^{(-1)}(z) & = \alpha(z) S(z)  \  ,  \label{solute:expansion:1}  \\
C_I^{(0)}(z)  & =
  C_{\infty} +  \alpha(z) S(z) \log \left( \frac{2}{S(z)} \right) - \frac{1}{2} \int_{-1}^1 \frac{\mathrm{d}}{\mathrm{d} \xi } \Big[\alpha({\xi})S({\xi})\Big] \ \mbox{sgn}\left( z - {\xi} \right) \log|z-{\xi}| \ d{\xi}  \  .
 \label{solute:expansion:0}
\end{align}  \\
\par \emph{(Velocity)}:
\tl{We first match the $\mathcal{O}(\epsilon, \lambda^0)$ velocities between the 
inner (region II) with the intermediate (region I)} 
\begin{equation}
\mbox{(II)} \quad \lim_{\lambda \rightarrow 0;  \ \rho \rightarrow \infty}  \epsilon \,  \bs{\mathcal{V}}^{(0)}  \   =  \bs {V}^{\text{slip}}(z;\epsilon) \quad  \qquad \mbox{(I)}  \ .
\end{equation}
\tl{Next we match the $\mathcal{O}(\epsilon, \lambda^0)$ velocities between the 
outer (region 0) with the intermediate (region I)} 
\begin{equation}
\mbox{(I)} \quad  V_{I,z}(R,z;\epsilon) =  V_{z}^{\text{slip}}   + \mathcal{W}(z;\epsilon) \ \log{\left( \frac{R}{S(z)} \right)} = \lim_{r  \rightarrow  \epsilon R  \ll 1} V_{0,z}(r,z) \quad  \qquad \mbox{(0)}  \  ,     \label{vel:matching:cyl:bulk}
\end{equation}
\tl{where $V_{z}^{\text{slip}} = \bs{V}_{}^{\text{slip}} \cdot \hat{\bs{e}}_z$. From equations (\ref{flow:approx:region:0})  and (\ref{flow:field:intermediate})  we find the unknown $\mathcal{W}(z;\epsilon)$ and $q(z)$,  independent of the choice of $R$, as long as $\epsilon R \ll 1$~\cite{cox1970motion,Batchelor70Slender,keller1976slender}.
Thus, $\mathcal{W} (z;\epsilon) = - 4 q (z;\epsilon)$ and}
\begin{equation}
U +   \left( \hat{\bs{e}}_z \cdot \hat{\bs{t}}(z) \right) V^{\text{slip}} \ =  4 q(z;\epsilon)  \log{\left( \frac{1}{\epsilon} \right)} + 4 q(z;\epsilon) \log\left( \frac{2 \sqrt{1-z^2} }{S(z)} \right) - 2q(z;\epsilon) + 2 \int_{-1}^{1}  \frac{q({\xi};\epsilon)  - q(z;\epsilon)}{|z - {\xi}|} d{\xi} \  , \label{speed:axial:matching}
\end{equation}
\tl{where , $V_{}^{\text{slip}} = \left | \bs{V}_{}^{\text{slip}}\right|$ and $  \hat{\bs{t}} (z) = \hat{\bs{t}} (S(z)) $.} \tll{ The propulsion speed $U$ is determined by requiring (see Section \ref{propulsion:speed:calcs}) that the solution of the  integral equation above satisfies the constraint of zero total force on the swimmer, i.e requiring}  
\beq \int_{-1}^{1}q(z)dz=0 \quad . \nonumber \label{eq:zero_force}
\eeq
\bibliography{Complete_Reference,library,library_1nov17}

\providecommand{\noopsort}[1]{}\providecommand{\singleletter}[1]{#1}%
\begin{thebibliography}{34}%
\makeatletter
\providecommand \@ifxundefined [1]{%
 \@ifx{#1\undefined}
}%
\providecommand \@ifnum [1]{%
 \ifnum #1\expandafter \@firstoftwo
 \else \expandafter \@secondoftwo
 \fi
}%
\providecommand \@ifx [1]{%
 \ifx #1\expandafter \@firstoftwo
 \else \expandafter \@secondoftwo
 \fi
}%
\providecommand \natexlab [1]{#1}%
\providecommand \enquote  [1]{``#1''}%
\providecommand \bibnamefont  [1]{#1}%
\providecommand \bibfnamefont [1]{#1}%
\providecommand \citenamefont [1]{#1}%
\providecommand \href@noop [0]{\@secondoftwo}%
\providecommand \href [0]{\begingroup \@sanitize@url \@href}%
\providecommand \@href[1]{\@@startlink{#1}\@@href}%
\providecommand \@@href[1]{\endgroup#1\@@endlink}%
\providecommand \@sanitize@url [0]{\catcode `\\12\catcode `\$12\catcode
  `\&12\catcode `\#12\catcode `\^12\catcode `\_12\catcode `\%12\relax}%
\providecommand \@@startlink[1]{}%
\providecommand \@@endlink[0]{}%
\providecommand \url  [0]{\begingroup\@sanitize@url \@url }%
\providecommand \@url [1]{\endgroup\@href {#1}{\urlprefix }}%
\providecommand \urlprefix  [0]{URL }%
\providecommand \Eprint [0]{\href }%
\providecommand \doibase [0]{http://dx.doi.org/}%
\providecommand \selectlanguage [0]{\@gobble}%
\providecommand \bibinfo  [0]{\@secondoftwo}%
\providecommand \bibfield  [0]{\@secondoftwo}%
\providecommand \translation [1]{[#1]}%
\providecommand \BibitemOpen [0]{}%
\providecommand \bibitemStop [0]{}%
\providecommand \bibitemNoStop [0]{.\EOS\space}%
\providecommand \EOS [0]{\spacefactor3000\relax}%
\providecommand \BibitemShut  [1]{\csname bibitem#1\endcsname}%
\let\auto@bib@innerbib\@empty
\bibitem [{\citenamefont {Marchetti}\ \emph {et~al.}(2013)\citenamefont
  {Marchetti}, \citenamefont {Joanny}, \citenamefont {Ramaswamy}, \citenamefont
  {Liverpool}, \citenamefont {Prost}, \citenamefont {Rao},\ and\ \citenamefont
  {Simha}}]{Marchetti2013}%
  \BibitemOpen
  \bibfield  {author} {\bibinfo {author} {\bibfnamefont {M.~C.}\ \bibnamefont
  {Marchetti}}, \bibinfo {author} {\bibfnamefont {J.~F.}\ \bibnamefont
  {Joanny}}, \bibinfo {author} {\bibfnamefont {S.}~\bibnamefont {Ramaswamy}},
  \bibinfo {author} {\bibfnamefont {T.~B.}\ \bibnamefont {Liverpool}}, \bibinfo
  {author} {\bibfnamefont {J.}~\bibnamefont {Prost}}, \bibinfo {author}
  {\bibfnamefont {M.}~\bibnamefont {Rao}}, \ and\ \bibinfo {author}
  {\bibfnamefont {R.~A.}\ \bibnamefont {Simha}},\ }\bibfield  {title} {\enquote
  {\bibinfo {title} {{Hydrodynamics of soft active matter}},}\ }\href {\doibase
  10.1103/RevModPhys.85.1143} {\bibfield  {journal} {\bibinfo  {journal} {Rev.
  Mod. Phys.}\ }\textbf {\bibinfo {volume} {85}},\ \bibinfo {pages}
  {1143--1189} (\bibinfo {year} {2013})}\BibitemShut {NoStop}%
\bibitem [{\citenamefont {Ramaswamy}(2010)}]{ramaswamy2010}%
  \BibitemOpen
  \bibfield  {author} {\bibinfo {author} {\bibfnamefont {S.}~\bibnamefont
  {Ramaswamy}},\ }\bibfield  {title} {\enquote {\bibinfo {title} {{The
  Mechanics and Statistics of Active Matter}},}\ }\href {\doibase
  10.1146/annurev-conmatphys-070909-104101} {\bibfield  {journal} {\bibinfo
  {journal} {Annu. Rev. Condens. Matter Phys.}\ }\textbf {\bibinfo {volume}
  {1}},\ \bibinfo {pages} {323--345} (\bibinfo {year} {2010})}\BibitemShut
  {NoStop}%
\bibitem [{\citenamefont {Julicher}\ \emph {et~al.}(2007)\citenamefont
  {Julicher}, \citenamefont {Kruse}, \citenamefont {Prost},\ and\ \citenamefont
  {Joanny}}]{Julicher2007}%
  \BibitemOpen
  \bibfield  {author} {\bibinfo {author} {\bibfnamefont {F.}~\bibnamefont
  {Julicher}}, \bibinfo {author} {\bibfnamefont {K.}~\bibnamefont {Kruse}},
  \bibinfo {author} {\bibfnamefont {J.}~\bibnamefont {Prost}}, \ and\ \bibinfo
  {author} {\bibfnamefont {J.}~\bibnamefont {Joanny}},\ }\bibfield  {title}
  {\enquote {\bibinfo {title} {{Active behavior of the Cytoskeleton}},}\ }\href
  {\doibase 10.1016/j.physrep.2007.02.018} {\bibfield  {journal} {\bibinfo
  {journal} {Phys. Rep.}\ }\textbf {\bibinfo {volume} {449}},\ \bibinfo {pages}
  {3--28} (\bibinfo {year} {2007})}\BibitemShut {NoStop}%
\bibitem [{\citenamefont {Golestanian}, \citenamefont {Liverpool},\ and\
  \citenamefont {Ajdari}(2005)}]{golestanian2005propulsion}%
  \BibitemOpen
  \bibfield  {author} {\bibinfo {author} {\bibfnamefont {R.}~\bibnamefont
  {Golestanian}}, \bibinfo {author} {\bibfnamefont {T.~B.}\ \bibnamefont
  {Liverpool}}, \ and\ \bibinfo {author} {\bibfnamefont {A.}~\bibnamefont
  {Ajdari}},\ }\bibfield  {title} {\enquote {\bibinfo {title} {Propulsion of a
  molecular machine by asymmetric distribution of reaction products},}\
  }\href@noop {} {\bibfield  {journal} {\bibinfo  {journal} {Physical review
  letters}\ }\textbf {\bibinfo {volume} {94}},\ \bibinfo {pages} {220801}
  (\bibinfo {year} {2005})}\BibitemShut {NoStop}%
\bibitem [{\citenamefont {Yariv}(2010)}]{yariv2010electrokinetic}%
  \BibitemOpen
  \bibfield  {author} {\bibinfo {author} {\bibfnamefont {E.}~\bibnamefont
  {Yariv}},\ }\bibfield  {title} {\enquote {\bibinfo {title} {Electrokinetic
  self-propulsion by inhomogeneous surface kinetics},}\ }in\ \href@noop {}
  {\emph {\bibinfo {booktitle} {Proceedings of the Royal Society of London A:
  Mathematical, Physical and Engineering Sciences}}}\ (\bibinfo {organization}
  {The Royal Society},\ \bibinfo {year} {2010})\ p.\ \bibinfo {pages}
  {rspa20100503}\BibitemShut {NoStop}%
\bibitem [{\citenamefont {Michelin}\ and\ \citenamefont
  {Lauga}(2014)}]{michelin2014phoretic}%
  \BibitemOpen
  \bibfield  {author} {\bibinfo {author} {\bibfnamefont {S.}~\bibnamefont
  {Michelin}}\ and\ \bibinfo {author} {\bibfnamefont {E.}~\bibnamefont
  {Lauga}},\ }\bibfield  {title} {\enquote {\bibinfo {title} {Phoretic
  self-propulsion at finite p{\'e}clet numbers},}\ }\href@noop {} {\bibfield
  {journal} {\bibinfo  {journal} {Journal of Fluid Mechanics}\ }\textbf
  {\bibinfo {volume} {747}},\ \bibinfo {pages} {572--604} (\bibinfo {year}
  {2014})}\BibitemShut {NoStop}%
\bibitem [{\citenamefont {Paxton}, \citenamefont {Sen},\ and\ \citenamefont
  {Mallouk}(2005)}]{paxton2005motility}%
  \BibitemOpen
  \bibfield  {author} {\bibinfo {author} {\bibfnamefont {W.~F.}\ \bibnamefont
  {Paxton}}, \bibinfo {author} {\bibfnamefont {A.}~\bibnamefont {Sen}}, \ and\
  \bibinfo {author} {\bibfnamefont {T.~E.}\ \bibnamefont {Mallouk}},\
  }\bibfield  {title} {\enquote {\bibinfo {title} {Motility of catalytic
  nanoparticles through self-generated forces},}\ }\href@noop {} {\bibfield
  {journal} {\bibinfo  {journal} {Chemistry--A European Journal}\ }\textbf
  {\bibinfo {volume} {11}},\ \bibinfo {pages} {6462--6470} (\bibinfo {year}
  {2005})}\BibitemShut {NoStop}%
\bibitem [{\citenamefont {Yariv}\ and\ \citenamefont
  {Schnitzer}(2013)}]{yariv2013electrophoretic}%
  \BibitemOpen
  \bibfield  {author} {\bibinfo {author} {\bibfnamefont {E.}~\bibnamefont
  {Yariv}}\ and\ \bibinfo {author} {\bibfnamefont {O.}~\bibnamefont
  {Schnitzer}},\ }\bibfield  {title} {\enquote {\bibinfo {title} {The
  electrophoretic mobility of rod-like particles},}\ }\href@noop {} {\bibfield
  {journal} {\bibinfo  {journal} {Journal of Fluid Mechanics}\ }\textbf
  {\bibinfo {volume} {719}},\ \bibinfo {pages} {R3} (\bibinfo {year}
  {2013})}\BibitemShut {NoStop}%
\bibitem [{\citenamefont {Ebbens}\ \emph {et~al.}(2014)\citenamefont {Ebbens},
  \citenamefont {Gregory}, \citenamefont {Dunderdale}, \citenamefont {Howse},
  \citenamefont {Ibrahim}, \citenamefont {Liverpool},\ and\ \citenamefont
  {Golestanian}}]{ebbens2014electrokinetic}%
  \BibitemOpen
  \bibfield  {author} {\bibinfo {author} {\bibfnamefont {S.}~\bibnamefont
  {Ebbens}}, \bibinfo {author} {\bibfnamefont {D.}~\bibnamefont {Gregory}},
  \bibinfo {author} {\bibfnamefont {G.}~\bibnamefont {Dunderdale}}, \bibinfo
  {author} {\bibfnamefont {J.}~\bibnamefont {Howse}}, \bibinfo {author}
  {\bibfnamefont {Y.}~\bibnamefont {Ibrahim}}, \bibinfo {author} {\bibfnamefont
  {T.}~\bibnamefont {Liverpool}}, \ and\ \bibinfo {author} {\bibfnamefont
  {R.}~\bibnamefont {Golestanian}},\ }\bibfield  {title} {\enquote {\bibinfo
  {title} {Electrokinetic effects in catalytic platinum-insulator janus
  swimmers},}\ }\href@noop {} {\bibfield  {journal} {\bibinfo  {journal} {EPL
  (Europhysics Letters)}\ }\textbf {\bibinfo {volume} {106}},\ \bibinfo {pages}
  {58003} (\bibinfo {year} {2014})}\BibitemShut {NoStop}%
\bibitem [{\citenamefont {Golestanian}, \citenamefont {Liverpool},\ and\
  \citenamefont {Ajdari}(2007)}]{golestanian2007designing}%
  \BibitemOpen
  \bibfield  {author} {\bibinfo {author} {\bibfnamefont {R.}~\bibnamefont
  {Golestanian}}, \bibinfo {author} {\bibfnamefont {T.}~\bibnamefont
  {Liverpool}}, \ and\ \bibinfo {author} {\bibfnamefont {A.}~\bibnamefont
  {Ajdari}},\ }\bibfield  {title} {\enquote {\bibinfo {title} {Designing
  phoretic micro-and nano-swimmers},}\ }\href@noop {} {\bibfield  {journal}
  {\bibinfo  {journal} {New Journal of Physics}\ }\textbf {\bibinfo {volume}
  {9}},\ \bibinfo {pages} {126} (\bibinfo {year} {2007})}\BibitemShut {NoStop}%
\bibitem [{\citenamefont {Anderson}(1989)}]{anderson01}%
  \BibitemOpen
  \bibfield  {author} {\bibinfo {author} {\bibfnamefont {J.~L.}\ \bibnamefont
  {Anderson}},\ }\bibfield  {title} {\enquote {\bibinfo {title} {Colloid
  transport by interfacial forces},}\ }\href@noop {} {\bibfield  {journal}
  {\bibinfo  {journal} {Annual Reviews of Fluid Mechanics}\ }\textbf {\bibinfo
  {volume} {21}},\ \bibinfo {pages} {61--99} (\bibinfo {year}
  {1989})}\BibitemShut {NoStop}%
\bibitem [{\citenamefont {Anderson}, \citenamefont {Lowell},\ and\
  \citenamefont {Prieve}(1982)}]{anderson1982motion}%
  \BibitemOpen
  \bibfield  {author} {\bibinfo {author} {\bibfnamefont {J.}~\bibnamefont
  {Anderson}}, \bibinfo {author} {\bibfnamefont {M.}~\bibnamefont {Lowell}}, \
  and\ \bibinfo {author} {\bibfnamefont {D.}~\bibnamefont {Prieve}},\
  }\bibfield  {title} {\enquote {\bibinfo {title} {Motion of a particle
  generated by chemical gradients part 1. non-electrolytes},}\ }\href@noop {}
  {\bibfield  {journal} {\bibinfo  {journal} {Journal of Fluid Mechanics}\
  }\textbf {\bibinfo {volume} {117}},\ \bibinfo {pages} {107--121} (\bibinfo
  {year} {1982})}\BibitemShut {NoStop}%
\bibitem [{\citenamefont {Paxton}\ \emph {et~al.}(2004)\citenamefont {Paxton},
  \citenamefont {Kistler}, \citenamefont {Olmeda}, \citenamefont {Sen},
  \citenamefont {St.~Angelo}, \citenamefont {Cao}, \citenamefont {Mallouk},
  \citenamefont {Lammert},\ and\ \citenamefont {Crespi}}]{paxton2004catalytic}%
  \BibitemOpen
  \bibfield  {author} {\bibinfo {author} {\bibfnamefont {W.~F.}\ \bibnamefont
  {Paxton}}, \bibinfo {author} {\bibfnamefont {K.~C.}\ \bibnamefont {Kistler}},
  \bibinfo {author} {\bibfnamefont {C.~C.}\ \bibnamefont {Olmeda}}, \bibinfo
  {author} {\bibfnamefont {A.}~\bibnamefont {Sen}}, \bibinfo {author}
  {\bibfnamefont {S.~K.}\ \bibnamefont {St.~Angelo}}, \bibinfo {author}
  {\bibfnamefont {Y.}~\bibnamefont {Cao}}, \bibinfo {author} {\bibfnamefont
  {T.~E.}\ \bibnamefont {Mallouk}}, \bibinfo {author} {\bibfnamefont {P.~E.}\
  \bibnamefont {Lammert}}, \ and\ \bibinfo {author} {\bibfnamefont {V.~H.}\
  \bibnamefont {Crespi}},\ }\bibfield  {title} {\enquote {\bibinfo {title}
  {Catalytic nanomotors: autonomous movement of striped nanorods},}\
  }\href@noop {} {\bibfield  {journal} {\bibinfo  {journal} {Journal of the
  American Chemical Society}\ }\textbf {\bibinfo {volume} {126}},\ \bibinfo
  {pages} {13424--13431} (\bibinfo {year} {2004})}\BibitemShut {NoStop}%
\bibitem [{\citenamefont {Moran}, \citenamefont {Wheat},\ and\ \citenamefont
  {Posner}(2010{\natexlab{a}})}]{posner01}%
  \BibitemOpen
  \bibfield  {author} {\bibinfo {author} {\bibfnamefont {J.~L.}\ \bibnamefont
  {Moran}}, \bibinfo {author} {\bibfnamefont {P.~M.}\ \bibnamefont {Wheat}}, \
  and\ \bibinfo {author} {\bibfnamefont {J.~D.}\ \bibnamefont {Posner}},\
  }\bibfield  {title} {\enquote {\bibinfo {title} {Locomotion of electrolytic
  nanomotors due to reaction induced charge autoelectrophoresis},}\ }\href@noop
  {} {\bibfield  {journal} {\bibinfo  {journal} {Physical Review E}\ }\textbf
  {\bibinfo {volume} {81}},\ \bibinfo {pages} {065302} (\bibinfo {year}
  {2010}{\natexlab{a}})}\BibitemShut {NoStop}%
\bibitem [{\citenamefont {Howse}\ \emph {et~al.}(2007)\citenamefont {Howse},
  \citenamefont {Jones}, \citenamefont {Ryan}, \citenamefont {Gough},
  \citenamefont {Vafabakhsh},\ and\ \citenamefont {Golestanian}}]{Howse2007}%
  \BibitemOpen
  \bibfield  {author} {\bibinfo {author} {\bibfnamefont {J.}~\bibnamefont
  {Howse}}, \bibinfo {author} {\bibfnamefont {R.}~\bibnamefont {Jones}},
  \bibinfo {author} {\bibfnamefont {A.}~\bibnamefont {Ryan}}, \bibinfo {author}
  {\bibfnamefont {T.}~\bibnamefont {Gough}}, \bibinfo {author} {\bibfnamefont
  {R.}~\bibnamefont {Vafabakhsh}}, \ and\ \bibinfo {author} {\bibfnamefont
  {R.}~\bibnamefont {Golestanian}},\ }\bibfield  {title} {\enquote {\bibinfo
  {title} {{Self-Motile Colloidal Particles: From Directed Propulsion to Random
  Walk}},}\ }\href {\doibase 10.1103/PhysRevLett.99.048102} {\bibfield
  {journal} {\bibinfo  {journal} {Phys. Rev. Lett.}\ }\textbf {\bibinfo
  {volume} {99}},\ \bibinfo {pages} {048102} (\bibinfo {year}
  {2007})}\BibitemShut {NoStop}%
\bibitem [{\citenamefont {Palacci}\ \emph {et~al.}(2013)\citenamefont
  {Palacci}, \citenamefont {Sacanna}, \citenamefont {Steinberg}, \citenamefont
  {Pine},\ and\ \citenamefont {Chaikin}}]{palacci2013living}%
  \BibitemOpen
  \bibfield  {author} {\bibinfo {author} {\bibfnamefont {J.}~\bibnamefont
  {Palacci}}, \bibinfo {author} {\bibfnamefont {S.}~\bibnamefont {Sacanna}},
  \bibinfo {author} {\bibfnamefont {A.~P.}\ \bibnamefont {Steinberg}}, \bibinfo
  {author} {\bibfnamefont {D.~J.}\ \bibnamefont {Pine}}, \ and\ \bibinfo
  {author} {\bibfnamefont {P.~M.}\ \bibnamefont {Chaikin}},\ }\bibfield
  {title} {\enquote {\bibinfo {title} {Living crystals of light-activated
  colloidal surfers},}\ }\href@noop {} {\bibfield  {journal} {\bibinfo
  {journal} {Science}\ }\textbf {\bibinfo {volume} {339}},\ \bibinfo {pages}
  {936--940} (\bibinfo {year} {2013})}\BibitemShut {NoStop}%
\bibitem [{\citenamefont {Yoshinaga}\ and\ \citenamefont
  {Liverpool}(2017)}]{Yoshinaga2017}%
  \BibitemOpen
  \bibfield  {author} {\bibinfo {author} {\bibfnamefont {N.}~\bibnamefont
  {Yoshinaga}}\ and\ \bibinfo {author} {\bibfnamefont {T.~B.}\ \bibnamefont
  {Liverpool}},\ }\bibfield  {title} {\enquote {\bibinfo {title} {{Hydrodynamic
  interactions in dense active suspensions : From polar order to dynamical
  clusters}},}\ }\href {\doibase 10.1103/PhysRevE.96.020603} {\bibfield
  {journal} {\bibinfo  {journal} {Phys. Rev. E - Stat. Nonlinear, Soft Matter
  Phys.}\ }\textbf {\bibinfo {volume} {96}},\ \bibinfo {pages} {020603(R)}
  (\bibinfo {year} {2017})}\BibitemShut {NoStop}%
\bibitem [{\citenamefont {Saha}, \citenamefont {Golestanian},\ and\
  \citenamefont {Ramaswamy}(2014)}]{Saha2014}%
  \BibitemOpen
  \bibfield  {author} {\bibinfo {author} {\bibfnamefont {S.}~\bibnamefont
  {Saha}}, \bibinfo {author} {\bibfnamefont {R.}~\bibnamefont {Golestanian}}, \
  and\ \bibinfo {author} {\bibfnamefont {S.}~\bibnamefont {Ramaswamy}},\
  }\bibfield  {title} {\enquote {\bibinfo {title} {{Clusters, asters, and
  collective oscillations in chemotactic colloids}},}\ }\href {\doibase
  10.1103/PhysRevE.89.062316} {\bibfield  {journal} {\bibinfo  {journal} {Phys.
  Rev. E - Stat. Nonlinear, Soft Matter Phys.}\ }\textbf {\bibinfo {volume}
  {89}},\ \bibinfo {pages} {1--7} (\bibinfo {year} {2014})},\ \Eprint
  {http://arxiv.org/abs/1309.4947} {arXiv:1309.4947} \BibitemShut {NoStop}%
\bibitem [{\citenamefont {Theurkauff}\ \emph {et~al.}(2012)\citenamefont
  {Theurkauff}, \citenamefont {Cottin-Bizonne}, \citenamefont {Palacci},
  \citenamefont {Ybert},\ and\ \citenamefont
  {Bocquet}}]{theurkauff2012dynamic}%
  \BibitemOpen
  \bibfield  {author} {\bibinfo {author} {\bibfnamefont {I.}~\bibnamefont
  {Theurkauff}}, \bibinfo {author} {\bibfnamefont {C.}~\bibnamefont
  {Cottin-Bizonne}}, \bibinfo {author} {\bibfnamefont {J.}~\bibnamefont
  {Palacci}}, \bibinfo {author} {\bibfnamefont {C.}~\bibnamefont {Ybert}}, \
  and\ \bibinfo {author} {\bibfnamefont {L.}~\bibnamefont {Bocquet}},\
  }\bibfield  {title} {\enquote {\bibinfo {title} {Dynamic clustering in active
  colloidal suspensions with chemical signaling},}\ }\href@noop {} {\bibfield
  {journal} {\bibinfo  {journal} {Physical review letters}\ }\textbf {\bibinfo
  {volume} {108}},\ \bibinfo {pages} {268303} (\bibinfo {year}
  {2012})}\BibitemShut {NoStop}%
\bibitem [{\citenamefont {Moran}, \citenamefont {Wheat},\ and\ \citenamefont
  {Posner}(2010{\natexlab{b}})}]{Moran2010}%
  \BibitemOpen
  \bibfield  {author} {\bibinfo {author} {\bibfnamefont {J.~L.}\ \bibnamefont
  {Moran}}, \bibinfo {author} {\bibfnamefont {P.~M.}\ \bibnamefont {Wheat}}, \
  and\ \bibinfo {author} {\bibfnamefont {J.~D.}\ \bibnamefont {Posner}},\
  }\bibfield  {title} {\enquote {\bibinfo {title} {{Locomotion of
  electrocatalytic nanomotors due to reaction induced charge
  autoelectrophoresis}},}\ }\href {\doibase 10.1103/PhysRevE.81.065302}
  {\bibfield  {journal} {\bibinfo  {journal} {Phys. Rev. E - Stat. Nonlinear,
  Soft Matter Phys.}\ }\textbf {\bibinfo {volume} {81}},\ \bibinfo {pages}
  {1--4} (\bibinfo {year} {2010}{\natexlab{b}})},\ \Eprint
  {http://arxiv.org/abs/1001.5462} {arXiv:1001.5462} \BibitemShut {NoStop}%
\bibitem [{\citenamefont {Popescu}\ \emph {et~al.}(2010)\citenamefont
  {Popescu}, \citenamefont {Dietrich}, \citenamefont {Tasinkevych},\ and\
  \citenamefont {Ralston}}]{popescu2010phoretic}%
  \BibitemOpen
  \bibfield  {author} {\bibinfo {author} {\bibfnamefont {M.~N.}\ \bibnamefont
  {Popescu}}, \bibinfo {author} {\bibfnamefont {S.}~\bibnamefont {Dietrich}},
  \bibinfo {author} {\bibfnamefont {M.}~\bibnamefont {Tasinkevych}}, \ and\
  \bibinfo {author} {\bibfnamefont {J.}~\bibnamefont {Ralston}},\ }\bibfield
  {title} {\enquote {\bibinfo {title} {Phoretic motion of spheroidal particles
  due to self-generated solute gradients},}\ }\href@noop {} {\bibfield
  {journal} {\bibinfo  {journal} {The European Physical Journal E}\ }\textbf
  {\bibinfo {volume} {31}},\ \bibinfo {pages} {351--367} (\bibinfo {year}
  {2010})}\BibitemShut {NoStop}%
\bibitem [{\citenamefont {Sabass}\ and\ \citenamefont
  {Seifert}(2012)}]{Sabass2012}%
  \BibitemOpen
  \bibfield  {author} {\bibinfo {author} {\bibfnamefont {B.}~\bibnamefont
  {Sabass}}\ and\ \bibinfo {author} {\bibfnamefont {U.}~\bibnamefont
  {Seifert}},\ }\bibfield  {title} {\enquote {\bibinfo {title} {{Nonlinear,
  electrocatalytic swimming in the presence of salt}},}\ }\href {\doibase
  10.1063/1.4719538} {\bibfield  {journal} {\bibinfo  {journal} {J. Chem.
  Phys.}\ }\textbf {\bibinfo {volume} {136}} (\bibinfo {year} {2012}),\
  10.1063/1.4719538},\ \Eprint {http://arxiv.org/abs/1202.3797}
  {arXiv:1202.3797} \BibitemShut {NoStop}%
\bibitem [{\citenamefont {Schnitzer}\ and\ \citenamefont
  {Yariv}(2015)}]{schnitzer2015osmotic}%
  \BibitemOpen
  \bibfield  {author} {\bibinfo {author} {\bibfnamefont {O.}~\bibnamefont
  {Schnitzer}}\ and\ \bibinfo {author} {\bibfnamefont {E.}~\bibnamefont
  {Yariv}},\ }\bibfield  {title} {\enquote {\bibinfo {title} {Osmotic
  self-propulsion of slender particles},}\ }\href@noop {} {\bibfield  {journal}
  {\bibinfo  {journal} {Physics of Fluids}\ }\textbf {\bibinfo {volume} {27}},\
  \bibinfo {pages} {031701} (\bibinfo {year} {2015})}\BibitemShut {NoStop}%
\bibitem [{\citenamefont {Jewell}, \citenamefont {Wang},\ and\ \citenamefont
  {Mallouk}(2016)}]{jewell2016catalytically}%
  \BibitemOpen
  \bibfield  {author} {\bibinfo {author} {\bibfnamefont {E.~L.}\ \bibnamefont
  {Jewell}}, \bibinfo {author} {\bibfnamefont {W.}~\bibnamefont {Wang}}, \ and\
  \bibinfo {author} {\bibfnamefont {T.~E.}\ \bibnamefont {Mallouk}},\
  }\bibfield  {title} {\enquote {\bibinfo {title} {Catalytically driven
  assembly of trisegmented metallic nanorods and polystyrene tracer
  particles},}\ }\href@noop {} {\bibfield  {journal} {\bibinfo  {journal} {Soft
  matter}\ }\textbf {\bibinfo {volume} {12}},\ \bibinfo {pages} {2501--2504}
  (\bibinfo {year} {2016})}\BibitemShut {NoStop}%
\bibitem [{\citenamefont {Nourhani}\ and\ \citenamefont
  {Lammert}(2016)}]{nourhani2016geometrical}%
  \BibitemOpen
  \bibfield  {author} {\bibinfo {author} {\bibfnamefont {A.}~\bibnamefont
  {Nourhani}}\ and\ \bibinfo {author} {\bibfnamefont {P.~E.}\ \bibnamefont
  {Lammert}},\ }\bibfield  {title} {\enquote {\bibinfo {title} {Geometrical
  performance of self-phoretic colloids and microswimmers},}\ }\href@noop {}
  {\bibfield  {journal} {\bibinfo  {journal} {Physical review letters}\
  }\textbf {\bibinfo {volume} {116}},\ \bibinfo {pages} {178302} (\bibinfo
  {year} {2016})}\BibitemShut {NoStop}%
\bibitem [{\citenamefont {Batchelor}(1970)}]{Batchelor70Slender}%
  \BibitemOpen
  \bibfield  {author} {\bibinfo {author} {\bibfnamefont {G.~K.}\ \bibnamefont
  {Batchelor}},\ }\bibfield  {title} {\enquote {\bibinfo {title} {Slender-body
  theory for particles of arbitrary cross-section in stokes flow},}\
  }\href@noop {} {\bibfield  {journal} {\bibinfo  {journal} {Journal of Fluid
  Mechanics}\ }\textbf {\bibinfo {volume} {44}},\ \bibinfo {pages} {419--440}
  (\bibinfo {year} {1970})}\BibitemShut {NoStop}%
\bibitem [{\citenamefont {Cox}(1970)}]{cox1970motion}%
  \BibitemOpen
  \bibfield  {author} {\bibinfo {author} {\bibfnamefont {R.~G.}\ \bibnamefont
  {Cox}},\ }\bibfield  {title} {\enquote {\bibinfo {title} {The motion of long
  slender bodies in a viscous fluid part 1. general theory},}\ }\href@noop {}
  {\bibfield  {journal} {\bibinfo  {journal} {Journal of Fluid mechanics}\
  }\textbf {\bibinfo {volume} {44}},\ \bibinfo {pages} {791--810} (\bibinfo
  {year} {1970})}\BibitemShut {NoStop}%
\bibitem [{\citenamefont {Keller}\ and\ \citenamefont
  {Rubinow}(1976)}]{keller1976slender}%
  \BibitemOpen
  \bibfield  {author} {\bibinfo {author} {\bibfnamefont {J.~B.}\ \bibnamefont
  {Keller}}\ and\ \bibinfo {author} {\bibfnamefont {S.~I.}\ \bibnamefont
  {Rubinow}},\ }\bibfield  {title} {\enquote {\bibinfo {title} {Slender-body
  theory for slow viscous flow},}\ }\href@noop {} {\bibfield  {journal}
  {\bibinfo  {journal} {Journal of Fluid Mechanics}\ }\textbf {\bibinfo
  {volume} {75}},\ \bibinfo {pages} {705--714} (\bibinfo {year}
  {1976})}\BibitemShut {NoStop}%
\bibitem [{\citenamefont {Geer}(1976)}]{geer1976stokes}%
  \BibitemOpen
  \bibfield  {author} {\bibinfo {author} {\bibfnamefont {J.}~\bibnamefont
  {Geer}},\ }\bibfield  {title} {\enquote {\bibinfo {title} {Stokes flow past a
  slender body of revolution},}\ }\href@noop {} {\bibfield  {journal} {\bibinfo
   {journal} {Journal of Fluid Mechanics}\ }\textbf {\bibinfo {volume} {78}},\
  \bibinfo {pages} {577--600} (\bibinfo {year} {1976})}\BibitemShut {NoStop}%
\bibitem [{\citenamefont {Tuck}(1964)}]{tuck1964some}%
  \BibitemOpen
  \bibfield  {author} {\bibinfo {author} {\bibfnamefont {E.}~\bibnamefont
  {Tuck}},\ }\bibfield  {title} {\enquote {\bibinfo {title} {Some methods for
  flows past blunt slender bodies},}\ }\href@noop {} {\bibfield  {journal}
  {\bibinfo  {journal} {Journal of Fluid Mechanics}\ }\textbf {\bibinfo
  {volume} {18}},\ \bibinfo {pages} {619--635} (\bibinfo {year}
  {1964})}\BibitemShut {NoStop}%
\bibitem [{\citenamefont {Tillett}(1970)}]{Tillett1970}%
  \BibitemOpen
  \bibfield  {author} {\bibinfo {author} {\bibfnamefont {J.~P.~K.}\
  \bibnamefont {Tillett}},\ }\bibfield  {title} {\enquote {\bibinfo {title}
  {{Axial and transverse Stokes flow past slender axisymmetric bodies}},}\
  }\href@noop {} {\bibfield  {journal} {\bibinfo  {journal} {J. Fluid Mech.}\
  }\textbf {\bibinfo {volume} {44}},\ \bibinfo {pages} {401--417} (\bibinfo
  {year} {1970})}\BibitemShut {NoStop}%
\bibitem [{\citenamefont {Shklyaev}, \citenamefont {Brady},\ and\ \citenamefont
  {C{\'o}rdova-Figueroa}(2014)}]{shklyaev2014non}%
  \BibitemOpen
  \bibfield  {author} {\bibinfo {author} {\bibfnamefont {S.}~\bibnamefont
  {Shklyaev}}, \bibinfo {author} {\bibfnamefont {J.~F.}\ \bibnamefont {Brady}},
  \ and\ \bibinfo {author} {\bibfnamefont {U.~M.}\ \bibnamefont
  {C{\'o}rdova-Figueroa}},\ }\bibfield  {title} {\enquote {\bibinfo {title}
  {Non-spherical osmotic motor: chemical sailing},}\ }\href@noop {} {\bibfield
  {journal} {\bibinfo  {journal} {Journal of Fluid Mechanics}\ }\textbf
  {\bibinfo {volume} {748}},\ \bibinfo {pages} {488--520} (\bibinfo {year}
  {2014})}\BibitemShut {NoStop}%
\bibitem [{\citenamefont {Van~Dyke}(1975)}]{van1975perturbation}%
  \BibitemOpen
  \bibfield  {author} {\bibinfo {author} {\bibfnamefont {M.}~\bibnamefont
  {Van~Dyke}},\ }\href@noop {} {\emph {\bibinfo {title} {Perturbation methods
  in fluid mechanics}}}\ (\bibinfo  {publisher} {Parabolic Press,
  Incorporated},\ \bibinfo {year} {1975})\BibitemShut {NoStop}%
\bibitem [{\citenamefont {Yariv}(2011)}]{Yariv2011}%
  \BibitemOpen
  \bibfield  {author} {\bibinfo {author} {\bibfnamefont {E.}~\bibnamefont
  {Yariv}},\ }\bibfield  {title} {\enquote {\bibinfo {title} {{Electrokinetic
  self-propulsion by inhomogeneous surface kinetics}},}\ }\href {\doibase
  10.1098/rspa.2010.0503} {\bibfield  {journal} {\bibinfo  {journal} {Proc. R.
  Soc. A Math. Phys. Eng. Sci.}\ }\textbf {\bibinfo {volume} {467}},\ \bibinfo
  {pages} {1645--1664} (\bibinfo {year} {2011})}\BibitemShut {NoStop}%
\end{thebibliography}%

%

\end{document}